\documentclass[letterpaper, 10 pt, journal, romanappendices, onecolumn, twoside]{IEEEtran}

\usepackage{amsmath, amssymb}
\usepackage{graphicx}
\usepackage{dsfont}
\usepackage{stmaryrd}
\usepackage{mathrsfs}
\usepackage{subfigure}
\usepackage[usenames,dvipsnames]{color}
\usepackage{cite}
\usepackage{asymptote}
\usepackage{enumerate}
\usepackage[normalem]{ulem}

\DeclareMathAlphabet{\mpzc}{OT1}{pzc}{m}{it}

\newcommand{\mbf}{\mathbf}
\newcommand{\msf}{\mathsf}

\newcommand{\mbb}{\mathbb}
\newcommand{\mrm}{\mathrm}
\newcommand{\mcl}{\mathcal}

\newcommand{\mds}{\mathds}

\newcommand\ind{\protect\mathpalette{\protect\independent}{\perp}}
\def\independent#1#2{\mathrel{\rlap{$#1#2$}\mkern2mu{#1#2}}}

\def\p#1{p_{\mathsf{#1}}}

\newtheorem{thm}{Theorem}
\newtheorem{lem}[thm]{Lemma}
\newtheorem{prop}[thm]{Proposition}
\newtheorem{cor}[thm]{Corollary}

\newtheorem{defn}[thm]{Definition}
\newtheorem{eg}[thm]{Example}
\newtheorem{rem}{Remark}
\newenvironment{dis}[1][Discussion :]{\begin{trivlist}
\item[\hskip \labelsep {\itshape #1} \hspace*{1mm}]}{\end{trivlist}}

\title{On the Capacity of Channels\\with {Timing} Synchronization Errors}
\author{Aravind~R.~Iyengar, Paul~H.~Siegel,~\IEEEmembership{Fellow,~IEEE} and Jack~K.~Wolf,~\IEEEmembership{Life~Fellow,~IEEE}
\thanks{A. R. Iyengar was with the Department of Electrical and Computer Engineering and the Center for Magnetic Recording Research, University of California, San Diego. He is now with Qualcomm Technologies Inc., Santa Clara CA 95051 USA (e-mail: ariyengar@qti.qualcomm.com). P. H. Siegel  is with the Department of Electrical and Computer Engineering and the Center for Magnetic Recording Research, University of California, San Diego, La Jolla, CA 92093 USA (e-mail: psiegel@ucsd.edu). J. K. Wolf (deceased) was with the Department of Electrical and Computer Engineering and the Center for Magnetic Recording Research, University of California, San Diego, La Jolla, CA 92093 USA.}
\thanks{This work was supported in part by the Center for Magnetic Recording Research and by the National Science Foundation under the Grant CCF-$0829865$. A summary of some of the results in Sections \ref{sec_sec} through \ref{sec_dord} was presented at the 2011 International Symposium on Information Theory (ISIT), St. Petersburg, Russia \cite{iye_11_isit_sec}.}}

\newcommand{\twobibs}[2]{#2}
\newcommand{\twofigs}[2]{#2}
\usepackage[colorlinks=true, pdfstartview=FitV, 
            linkcolor=black, citecolor=black, 
            urlcolor=black, plainpages=false,
            pdfpagelabels]{hyperref}

\begin{document}
\maketitle
\thispagestyle{plain}

\begin{abstract}
We consider a new formulation of a class of synchronization error channels and derive analytical bounds and numerical estimates for the capacity of these channels. For the binary channel with only deletions, we obtain an expression for the symmetric information rate in terms of subsequence weights which reduces to a tight lower bound for small deletion probabilities. We are also able to exactly characterize the Markov-$1$ rate for the binary channel with only replications. For a channel that introduces deletions as well as replications of input symbols, we design 
approximating channels 
{that} 
parameterize the state space 
and show that the information rates {of these approximate channels} approach that of the deletion-replication channel as the state space grows. For the case of the channel where deletions and replications occur with the same probabilities, a stronger result in the convergence of mutual information rates is shown. The numerous advantages this new formulation presents are explored.
\end{abstract}


\section{Introduction}

\IEEEPARstart{C}{hannels} with synchronization errors have been familiar to information and coding theorists and practitioners alike ever since the advent of the digital information era. Although Dobrushin \cite{dob_67_pit_sec} established the coding theorem for such channels as early as 1967, tackling these channels in terms of estimating information rates and constructing codes with good performance has proved to be very tough. In the last decade, significant progress has been made in estimating achievable information rates for certain channels with synchronization errors. However, a coding scheme with provably ``good'' performance remains elusive thus far.

In this paper, we start {by revisiting} 
Dobrushin's model of channels with synchronization errors, henceforth referred to as the \emph{synchronization error channel} (SEC). {We model channels with timing-synchronization errors as channels with states, and show that these channels are equivalent to a subclass of SECs considered by Dobrushin---called the deletion-replication channels (DRCs). 
Using our model for the DRC, we establish several bounds on achievable rates under special cases. We also construct a sequence of channels that ``approximate'' the DRC and whose limit is the DRC itself. We then use these approximate channels to derive some numerical estimates of the information rates achievable over the DRC.} Although the motivation behind the alternative model is straightforward, its use to obtain non-trivial bounds on the capacity of the SEC has, to the best of our knowledge, not been found in literature. While the present paper concerns only a few asymptotic results on information rates of {channels with timing synchronization errors}, we think that the model presented here can be utilized to design codes for SECs in general. 

The remainder of this paper is organized as follows. In Section \ref{sec_sec}, we revisit Dobrushin's model of a SEC and recall the main results on capacity of SECs. {
In 
Section \ref{sec_eqchn}
, we formulate our model for channels with timing synchronization errors and establish their equivalence to Dobrushin's model for the DRCs. We explore the advantages of our formulation of the DRC model in the two subsequent sections.} Under special cases of channels with only deletions or only replications, we give some simple, non-trivial and sometimes tight bounds on the capacity in Sections \ref{ssec_bdc} and \ref{ssec_bdupc}. We then construct a sequence of finite state channels that approximate the DRC and establish certain properties of this sequence of channels that serve as {estimates} 
for the capacity of the DRC in Section \ref{sec_drc}. 
We conclude with summary and remarks in Section \ref{sec_conc}.

\section{Synchronization Error Channels} \label{sec_sec}
\begin{rem}[Notation]
Non-random variables are written as lowercase letters, e.g., $n$. We denote sets by double-stroke uppercase letters, e.g., $\mbb{X}$. We will reserve $\mathbb{N}$, $\mathbb{Z}$ and $\mathbb{R}$ to denote the sets of natural numbers, integers and real numbers, respectively. $\mathbb{Z}^+$ denotes the set of non-negative integers. We define 
\begin{align}
[n] &\triangleq \{1, 2, \cdots, n\}, n \in \mathbb{N}, \\ 
[0] &\triangleq \emptyset, \\ 
[m : n] &\triangleq 
\begin{cases}
\{m, m + 1, \cdots, n\}, &m \leq n, \\
\emptyset, &n < m.
\end{cases} \text{ and }\\ 
\mathbb{Z}_{\pm m} &\triangleq \{-m, -m + 1, \cdots, 0, 1, \cdots, m\}{\ }\forall{\ }m \in \mathbb{Z}^+. 
\end{align}

For some $n \in \mathbb{N}$, we will let $\mathbb{X}^n$ denote the set of vectors of dimension $n$ with elements from $\mathbb{X}$. We will write $\overline{x}$ to denote a string, and $\lambda$ to denote the \emph{empty string}. The \emph{length} of a string, denoted $|\overline{x}|$, is the number of symbols in it, and by definition, $|\lambda| = 0$. With some abuse of notation, we will use ``vectors of dimension $n$'' and ``strings of length $n$'' interchageably. The set of all strings of length $n$ over the alphabet $\mathbb{X}$ is hence also denoted $\mathbb{X}^n$, and $\mathbb{X}^0 = \{\lambda\}$. We write $\overline{\mathbb{X}}$ to denote the set of all strings over the set $\mathbb{X}$, i.e.,
\begin{equation} 
\overline{\mathbb{X}} = \bigcup_{i = 0}^\infty \mathbb{X}^i.
\end{equation} 
{We will use the notation ``$\circ$'' to denote the concatenation operation, so that ${\overline{x}\circ\overline{y}}$ is the concatenation of strings $\overline{x}$ and $\overline{y}$. Similarly, $\circ(\overline{x}_{[n]})$ denotes concatenation of the $n$ strings $\overline{x}_i, i \in [n]$.}

Throughout the paper, we assume an underlying probability space $(\mathbb{S}, \mathscr{B}, \msf{P})$ over which random variables, denoted by uppercase letters, e.g., $X$, are defined. Random vectors are denoted by uppercase letters with the \emph{multiset} of indices as subscripts, e.g., $X_{[n]} = (X_1, X_2, \cdots, X_n)$, or $X_{Y_{[n]}}$ when the multiset of indices is itself the elements of a random vector $Y_{[n]}$. Random processes (assumed discrete-time) are denoted by script letters $\mcl{X}$, or subscripted by the set of natural numbers, $X_{\mathbb{N}}$.

We will use the asymptotic notations $O(\cdot)$, $o(\cdot)$, $\omega(\cdot)$ as in \cite{knu_76_sig_oot, cor_01_bok_algo}. {We will write $a_n \doteq b_n$ for real sequences $\{a_n\}_{n \geq 1}$ and $\{b_n\}_{n \geq 1}$ to mean }
${\lim_{n \rightarrow \infty} \frac{a_n}{n} = \lim_{n \rightarrow\infty} \frac{b_n}{n}}$.
\hfill$\square$\vspace{2mm}
\end{rem}

We start by defining the synchronization error channels as considered by Dobrushin \cite{dob_67_pit_sec}. \vspace{2mm}
\begin{defn}[Memoryless SECs] \label{def_msec}
Let $\mbb{X}$ be a finite set. A \emph{memoryless} synchronization error channel is specified by a stochastic matrix
\[
\{q(\overline{y}\mid x), \overline{y} \in \overline{\mbb{Y}}, x \in \mbb{X}\}
\]
where $\mbb{Y}$ is the output alphabet. From the properties of a stochastic matrix, we have
\begin{equation} \label{eq_condn1}
0 \leq q(\overline{y}\mid x) \leq 1,\qquad \sum_{\overline{y} \in \overline{\mbb{Y}}} q(\overline{y}\mid x) = 1{\ }\forall{\ }x \in \mbb{X}.
\end{equation}
Further, we will assume that the mean value of the length of the output string arising from one input symbol is strictly positive and finite, i.e.,
\begin{equation} \label{eq_condn2}
0 < \sum_{\overline{y} \in \overline{\mbb{Y}}} |\overline{y}|\cdot q(\overline{y}\mid x) < \infty.
\end{equation}
For $x_{[n]} = (x_1, x_2, \cdots, x_n) \in \mbb{X}^n$ and $\overline{y}_{[n]} = (\overline{y}_1, \overline{y}_2, \cdots, \overline{y}_n) \in \overline{\mbb{Y}}^n$, we write
\begin{equation} 
q_n(\overline{y}_{[n]}\mid x_{[n]}) = \prod_{i = 1}^n q(\overline{y}_i\mid x_i).
\end{equation} 
The transition probabilities of the memoryless SEC are defined as
\begin{equation} \label{eq_msectp}
Q_n(\overline{y}\mid x_{[n]}) = \sum_{{\circ(\overline{y}_{[n]})} = \overline{y}} q_n(\overline{y}_{[n]}\mid x_{[n]})
\end{equation}
for $\overline{y} \in \overline{\mbb{Y}}$ and $x_{[n]} \in \mbb{X}^n$. The memoryless SEC is given by the triplet $\mathbf{Q}_n \triangleq (\mathbb{X}, {\mathbb{Y},} Q_n)$, the input and the output alphabets, and the transition probabilities between input strings of length $n$ and all output strings.\hfill$\square$\vspace{2mm}
\end{defn}

Consider the sequence of memoryless SECs $\{\mathbf{Q}_n\}_{n = 1}^{\infty}$. Then, we have the following.
\vspace{2mm}
\begin{thm}[Capacity \cite{dob_67_pit_sec}] \label{thm_cap}
Let $X_{[n]}$ and $\overline{Y}$ denote the input and the output of the SEC $\mathbf{Q}_n$. Let
\begin{equation} 
C_n = \sup_{\mathsf{P}(X_{[n]})} \frac{1}{n}I(X_{[n]}; \overline{Y}).
\end{equation} 
Then,
\begin{equation} 
C = \lim_{n \rightarrow \infty} C_n = \inf_{n \geq 1} C_n
\end{equation} 
exists and is equal to the capacity of the sequence of SECs.\hfill$\blacksquare$\vspace{2mm}
\end{thm}
The quantity $C$ represents the maximum rate at which information can be transferred over the SEC with vanishing error probability. Furthermore, the following result shows that, in estimating the capacity of the SEC, we can restrict ourselves to a subclass of possible input processes $\mcl{X}$.\vspace{2mm}

\begin{prop}[Markov Capacity \cite{dob_67_pit_sec}] \label{prop_marcap}
Let $\mcl{X}_\mcl{M}$ be a stationary, ergodic, Markov process over $\mbb{X}$. Then the capacity of the sequence $\{\mathbf{Q}_n\}_{n = 1}^{\infty}$ is
\begin{equation} 
C = \sup_{\mcl{X}_\mcl{M}} \lim_{n \rightarrow \infty} \frac{1}{n}I(X_{[n]}; \overline{Y}).
\end{equation} 
The capacity is therefore the supremum of the rates achievable through stationary, ergodic, Markov processes $\mathcal{X}_{\mathcal{M}}$.
\hfill$\blacksquare$\vspace{2mm}
\end{prop}

We will now give an example of a memoryless SEC. Throughout the paper, we will assume that the input alphabet for the SECs is $\mathbb{X} = \{0, 1\}$, i.e., the channels considered are \emph{binary} memoryless SECs. However, we note here that all the results in the paper can be straightforwardly extended to the case where $\mathbb{X}$ is any finite set.\vspace{2mm}

\begin{eg}[Deletion-Replication Channel (DRC)] \label{eg_ddc}
Consider the binary SEC with $\mbb{X} = \mbb{Y} = \{0, 1\}$ and the following stochastic matrix.
\begin{equation} \label{eq_ddc1tp}
q(\overline{y}\mid x) =
\begin{cases}
p_{\msf{d}}, &\overline{y} = \lambda\\
p_{\msf{t}}p_{\msf{r}}^{\ell - 1}, &\overline{y} = x^\ell,{\ }\forall{\ }\ell \geq 1.
\end{cases}
\end{equation} 
Intuitively, we can think of $p_\msf{d}$ as the deletion probability, $p_\msf{t}$ as the transmission probability, and $p_\msf{r}$ as the replication probability, i.e., when $x \in \mbb{X}$ is sent, it is either deleted with probability $\p{d}$, or transmitted and replicated $(\ell - 1)$ times with probability $\p{t}\p{r}^{\ell - 1}$ for $\ell \geq 1$. From \eqref{eq_condn1}, we get for $p_\msf{r} < 1$
\begin{equation} 
p_{\msf{d}} + \sum_{\ell = 1}^{\infty} p_\msf{t}p_\msf{r}^{\ell - 1} = p_\msf{d} + \frac{p_\msf{t}}{1 - p_\msf{r}} = 1,
\end{equation} 
or equivalently
\begin{equation} \label{eq_trnprob}
p_\msf{t} = (1 - p_\msf{d})(1 - p_\msf{r}).
\end{equation}
From \eqref{eq_condn2},
\begin{equation} 
0 < \sum_{\ell = 1}^\infty \ell p_\msf{t}p_\msf{r}^{\ell - 1} = \frac{p_\msf{t}}{(1 - p_\msf{r})^2} = \frac{1 - p_\msf{d}}{1 - p_\msf{r}} < \infty
\end{equation} 
where we use Equation \eqref{eq_trnprob}. Hence $(p_\msf{d}, p_\msf{r}) \in [0, 1)^2$. Note that when $\p{r} = 0$, the DRC is the same as the \emph{binary deletion channel} (BDC); and when $\p{d} = 0$, it is the \emph{binary replication channel} (BRC), also referred to as the \emph{geometric binary sticky channel} \cite{mit_08_tit_sticky}.\hfill$\square$\vspace{2mm}
\end{eg}

%
\subsection{{Prior work}}
The BDC has been the most well-studied SEC. In \cite{mit_09_prs_delchn}, the author surveys the results that were known prior to 2009. To summarize, the best known lower bounds were obtained, chronologically, through bounds on the cutoff rate for sequential decoding \cite{gal_61_llr_seqdecsec}, bounding the rate with a first-order Markov input \cite{dig_06_tit_finbufchn}, reduction to a Poisson-repeat channel \cite{mit_06_tit_lbcapdc}, analyzing a ``jigsaw-puzzle'' coding scheme \cite{dri_07_tit_implbddc}, or by directly bounding the information rate by analyzing the channel as a joint renewal process \cite{kir_10_tit_ddccap}. Recently, \cite{kan_10_isit_delchn} and \cite{kal_10_isit_delchn} independently gave the capacity of a BDC with small deletion probabilities, and showed that it is achieved by independent and uniformly distributed (i.u.d.) inputs. The known upper bounds for the BDC have been obtained by genie-aided decoder arguments \cite{dig_07_isit_capubdc, fer_10_tit_bdccap}. An idea from \cite{fer_10_tit_bdccap} was extended to obtain some analytical lower bounds on the capacity of channels that involve substitution errors as well as insertions or deletions \cite{rah_11_arx_capid}. The idea in \cite{kan_10_isit_delchn} was extended to obtain a better approximation for the capacity of the BDC with small deletion probabilities in \cite{kan_13_tit_bdc}. {More recently, the authors of \cite{mer_12_tit_nsc} obtained numerical lower bounds on the capacity of the BDC. To do this, they estimate the information rates achieved by Markov sources through a Monte-Carlo estimation of the output entropy rate and an estimation of the conditional output entropy (conditioned on the input process) via a sum-product algorithm; and then optimized the input distribution using the Nelder-Mead algorithm.}

For the BRC, \cite{mit_08_tit_sticky} obtained lower bounds on the capacity by numerically estimating the capacity per unit cost of the equivalent channel of runs through optimization of $8$ and $16$ bit codes. {This approach was further improved in \cite{mer_12_tit_nsc} to obtain sharp 
lower and upper bounds (with negligible gap). These results showed a surprising characteristic of the BRC---that the capacity approaches a non-zero value as the probability of replication approaches $1$.}

{There are very few results on channels including both deletions and replications. Although the method in \cite{kir_10_tit_ddccap} can be applied to such channels, no estimates of the achievable rates were presented therein. In \cite{mer_12_tit_nsc}, the authors estimate numerical lower bounds on the capacity of binary DRC using the same approach as for BDCs mentioned above. They also present upper bounds on the capacity by estimating the capacity per unit-cost of an equivalent block channel with a genie-aided decoder that truncates input and output runs.}

%
\subsection{{Contributions}}
{In contrast to the existing results on the BDC}, our approach explicitly characterizes the achievable information rates in terms of ``subsequence-weights'', which is a measure relevant in maximum-likelihood (ML) decoding for the BDC \cite{mit_09_prs_delchn}. Additionally, the method proposed here gives the tight bound on capacity for small deletion probabilities obtained in \cite{kan_10_isit_delchn} more directly\footnote{Note that although we obtain the same lower bound for the capacity of the BDC as in \cite{kan_10_isit_delchn}, we do not prove a converse here.}. {In contrast to the 
lower bound in \cite{mer_12_tit_nsc}, our results are analytical and readily generalize to any finite alphabet.}

{For the BRC,} we obtain direct analytical lower bounds on the capacity, 
including an exact expression for the Markov-$1$ rate {\textcolor{black}{that was evaluating numerically in} 
\cite{mer_12_tit_nsc}. Interestingly, the expression for the Markov-$1$ rate presented in \cite{dri_07_tit_implbddc} is different from our result, although the values of the expressions match.}

{For the case of DRCs, by approximating them using finite-state channels (FSCs), we establish 
results that show that the information rates of these approximating channels approach that of the DRC as the state space grows. For the case of DRCs with equal deletion and replication probabilities, we prove that a sublinear growth in the state space of the approximating FSCs is sufficient for the convergence of information rates to that of the underlying DRC. We also estimate numerically the information rates achievable on these FSCs. \textcolor{black}{Although these estimates are not direct bounds on the capacity of the DRC,} 
we provide empirical evidence to support the claim that the information rates of these approximating FSCs converge a lot more quickly than is guaranteed by theory for small deletion-replication probabilities. This implies that in the range of small deletion-replication probabilities, our estimates serve as good upper bounds for the capacity of the DRC. Moreover, due to an artifact of the channel model considered in \cite{mer_12_tit_nsc}, the complete channel space is not representable, and therefore, for the channels that cannot be represented by the model, no results are known. Our approach does not have this limitation and the results we obtain apply to all channels in the family of DRCs.}


\section{{Timing Synchronization Error Channels}}
 \label{sec_eqchn}
We now {present a mathematical model for channels with timing synchronization errors. Intuitively, the noise in timing synchronization error channels manifests as a time-drift between the inputs and the outputs of the channel.}

\subsection{Channel Model} \label{ssec_chnmod}
\textcolor{black}{We model the timing synchronization error channels as channels with states where the state governs the \emph{drift} between the channel input and output. The realization of the channel state process produces a noisy version of the channel input at its output where the exact correspondence between the output symbol \emph{indices} and the input symbol indices are lost due to randomness. We start with a few definitions that will help us in describing the channel model that will be considered throughout the paper.}
\begin{defn}[State Process] \label{def_state}
{The \emph{state process} $\mathcal{Z}$ is defined to be a first-order Markov process over the set of integers $\mathbb{Z}$} 
\textcolor{black}{with the transition probabilities, for $i \in \mathbb{Z}$, given by}
\begin{equation} \label{eq_ztp}
{\msf{P}(Z_i = z_i \mid Z_{i - 1} = z_{i - 1}) = \begin{cases} \p{r}, &z_i = z_{i - 1} + 1\\ \p{d}^{\ell}\p{t}, &z_i = z_{i - 1} - \ell{\ }\forall{\ }\ell \geq 0\end{cases}}
\end{equation}
\textcolor{black}{where }{$(\p{d}, \p{r}) \in [0, 1)^2$ and $\p{t}$ is as defined in Equation \eqref{eq_trnprob}.\hfill$\square$\vspace{2mm}
}
\end{defn}
\textcolor{black}{We note that the transition probabilities defined above are independent of the time index $i$. Also, the probabilities are only functions of the ``increments'' of the state process $(z_i - z_{i - 1})$.}

\begin{defn}[Compatible State Paths] \label{def_comppath}
\textcolor{black}{For $n \in \mathbb{N}$,}{ a state \emph{path} $z_{[n]} \in \mathbb{Z}^n$ that has a strictly positive probability according to the} \textcolor{black}{ transition probabilities in Equation 
\eqref{eq_ztp}}{ is called \emph{compatible}.
\hfill$\square$\vspace{2mm}
}
\end{defn}
\textcolor{black}{Therefore, without loss of generality, we will confine attention to the full set comprising only compatible state paths in our discussion throughout the paper.}

\begin{defn}[Index Process] \label{def_index}
{We define the \emph{index process} $\varGamma$ as the random process related to the state process $\mathcal{Z}$ as}
\begin{align}
{}\Gamma_i \triangleq i - Z_i{\ }\forall{\ }i \textcolor{black}{\in \mathbb{Z}.}\tag*{$\square$}
\end{align}
\end{defn}
\textcolor{black}{It is easy to see that the index process inherits the first-order Markovity from the $\mathcal{Z}$ process. Moreover, we also have that the $\varGamma$ process is almost surely non-decreasing, i.e.,}
\begin{equation} 
\textcolor{black}{\Gamma_{i + j} \geq \Gamma_i{\ }\forall{\ }i \in \mathbb{Z}, j \geq 0{\ }\text{a. s.},}
\end{equation} 
\textcolor{black}{a fact that we will use often throughout the paper.}

\begin{defn}[Maximal Index Times] \label{def_idxtime}
{For} $n \textcolor{black}{\in \mathbb{Z}}$, {we define the $n^{\text{th}}$ \emph{maximal index time}, denoted $N_n$, as}
\begin{equation}
{N_n \triangleq \sup \{}\textcolor{black}{i \in \mathbb{Z}} : {\Gamma_i \leq n\}.}
\end{equation}
{In words, the $n^{\text{th}}$ maximal index time is the maximum time instant for which the index process is bounded by $n$.\hfill$\square$\vspace{2mm}}
\end{defn}
\textcolor{black}{Although we have thus far defined the state and index processes to be doubly-infinite random processes, we will henceforth be only interested in the semi-infinite processes $\{Z_n\}_{n\in\mbb{Z}^+}$ and $\{\Gamma_n\}_{n\in\mbb{Z}^+}$. We further impose the following boundary conditions which we will later expound}
\begin{equation} \label{eq_boundary_condn}
\textcolor{black}{{\mathfrak{B}_0} \triangleq \{Z_0 = 0 \text{ and } N_0 = 0\}.}
\end{equation}
\textcolor{black}{With the above boundary conditions, it is easy to see that $Z_{[N_n]} \leftrightarrow \Gamma_{[N_n]} \leftrightarrow N_{[n]}$ for every $n \in \mathbb{N}$.}
\begin{defn}[Index Vector and Index Set] \label{def_idxvecset}
{For $n \in \mathbb{N}$, the random vector $\Gamma_{[N_n]}$ is referred to as the \emph{index vector}. We denote by $\{\Gamma_{[N_n]}\}$ the set of elements in the index vector $\Gamma_{[N_n]}$ and refer to it as the \emph{index set}.\hfill$\square$\vspace{2mm}}
\end{defn}

\begin{figure*}[!b]
\hrule
\begin{align}
P_n&(\overline{y} \mid x_{[n]}, \textcolor{black}{\mathfrak{B}_0}) \notag \\
&= 
\textcolor{black}{\sum_{\{z_{[|\overline{y}|]} : z_1 < 1\}} \msf{P}\Big(N_n = |\overline{y}|, Z_{[|\overline{y}|]} = z_{[|\overline{y}|]}, Y_{[|\overline{y}|]} = \overline{y} \mid X_{[n]} = x_{[n]}, \mathfrak{B}_0\Big)} \\ 
&{= \sum_{\{z_{[|\overline{y}|]} : z_1 < 1\}} \msf{P}(Z_{|\overline{y}| + 1} < |\overline{y}| + 1 - n \mid Z_{|\overline{y}|} = z_{|\overline{y}|})}\cdot\textcolor{black}{\msf{P}(Z_1 = z_1 \mid \mathfrak{B}_0)}\cdot\msf{P}(Y_{1} = y_1 \mid X_{[n]} = x_{[n]}, Z_{1} = z_1) \notag \\
&\qquad\qquad\qquad\qquad{\cdot\prod_{i = 2}^{|\overline{y}|}\Big(\msf{P}(Z_{i} = z_i \mid Z_{i - 1} = z_{i - 1})\cdot\msf{P}(Y_{i} = y_i \mid X_{[n]} = x_{[n]}, Z_{i} = z_i)\Big)} \\
&{= \textcolor{black}{\frac{1}{1 - \p{r}}}\cdot\sum_{\{z_{[|\overline{y}|]} : z_1 < 1\}} \msf{P}(Z_{|\overline{y}| + 1} < |\overline{y}| + 1 - n \mid Z_{|\overline{y}|} = z_{|\overline{y}|})\cdot\prod_{i = 1}^{|\overline{y}|}\Big(\msf{P}(Z_{i} = z_i \mid Z_{i - 1} = z_{i - 1})\cdot\mds{1}_{\{y_i = x_{i - z_i}\}}\Big)} \label{eq_eqchntrnp} \\
&{= \sum_{\{z_{[|\overline{y}|]}:z_1<1\}}\p{d}^{n - |\overline{y}| + z_{|\overline{y}|}}\cdot\prod_{i = 1}^{|\overline{y}|}\Big(\msf{P}(Z_i = z_i\mid Z_{i - 1} = z_{i - 1})\mds{1}_{\{y_i = x_{i - z_i}\}}\Big).} \label{eq_eqchntp}
\end{align}
\end{figure*}
We can now define the {channel model for timing synchronization errors.}
\begin{defn}[{Channel Model for Timing Synchronization Errors}]
\label{def_ddcstate}
{Let $\mathbb{X}$ be a finite set. For a fixed $n \in \mathbb{N}$, the channel input is a realization $X_{[n]} \in \mathbb{X}^n$ of the input process $\mathcal{X}$. We define the channel output as}
\begin{equation} \label{eq_eqchnmod}
Y_i = X_{\Gamma_i} = X_{i - Z_i},{\ }\forall{\ }i \in [\textcolor{black}{N_n}],
\end{equation}
where the state process $\mcl{Z}$ is assumed to be independent of the channel input process {$\mathcal{X}$, i.e., $\mathcal{X} \ind \mathcal{Z}$.}\hfill$\square$\vspace{2mm}
\end{defn}
\begin{rem}
{We will henceforth assume that $\mathbb{S} = \mathbb{S}_\mathcal{X} \times \mathbb{S}_\mathcal{Z}$ and that $\mathscr{B} = \mathscr{B}_\mathcal{X} \times \mathscr{B}_\mathcal{Z}$ with $\mathscr{B}_\mathcal{X} = \sigma(\mathcal{X})$ and $\mathscr{B}_\mathcal{Z} = \sigma(\mathcal{Z})$, i.e., the space $(\mathbb{S}, \mathscr{B})$ is a product space. Since in our channel model in Definition \ref{def_ddcstate}, we have $\mathcal{X} \ind \mathcal{Z}$, there is no loss of generality in making this assumption on the space $(\mathbb{S}, \mathscr{B})$ where the probability measure $\msf{P}$ is defined.\vspace*{2mm}}
\end{rem}

\textcolor{black}{It is clear that the channel output alphabet $\mathbb{Y} = \mathbb{X}$. With the model above, we can now provide an operational significance of the boundary condition $\mathfrak{B}_0$ in Equation \eqref{eq_boundary_condn}. $Z_0 = \Gamma_0 = 0$ implies that the channel is perfectly synchronized before transmission commences, and $N_0 = 0$ implies that the channel commences transmission of input symbols of interest, i.e., input symbols with indices in the set $[n]$, at time index $1$. The random variable $N_n$ can also be interpreted as the length of the output produced by the channel upon feeding an input of length $n$.}

For $\overline{y} \in \overline{\mbb{Y}}$ and $x_{[n]} \in \mbb{X}^n$, the channel transition probabilities are given as in Equation \eqref{eq_eqchntp}{, where we have assumed that $z_0 = 0$}\textcolor{black}{, and \eqref{eq_eqchntp} follows from \eqref{eq_eqchntrnp} from the transition probabilities \eqref{eq_ztp} and the fact that $\gamma_{|\overline{y}|} = |\overline{y}| -z_{|\overline{y}|} \leq n$}. 
\textcolor{black}{We emphasize that the channel transition probabilities $P_n$ are implicitly conditioned upon the boundary conditions $\mathfrak{B}_0$.} 
\begin{defn}[Timing Synchronization Error Channels] \label{def_tsec}
For each $n \in \mathbb{N}$, we define the {timing synchronization error channel} 
as the channel $\mathbf{P}_n \triangleq (\mathbb{X}{, \mathbb{Y}}, P_n)$ where $\mbb{Y} = \mbb{X}$ and $P_n$ is as in Equation \eqref{eq_eqchntp}.\hfill$\square$\vspace{2mm}
\end{defn}

\begin{figure}[!ht]
\centering
\twofigs{\include{zpaths}}{\includegraphics{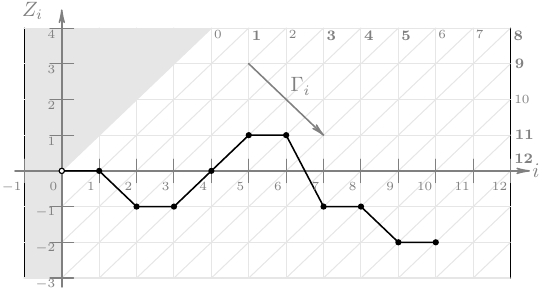}}
\caption{{Plot of state path realization $z_{[n]}$ for $n = 10$. The corresponding realization of the $\varGamma$ process can be seen by drawing contours of $\Gamma_i$: $Z_i = i - \gamma$. These contours are also shown in the figure, and the corresponding $\Gamma_i$ values are noted along the top and the right edges of the plot. The $\Gamma_{[10]}$ values realized by the state path $z_{[10]}$ are shown in bold larger font. 
The space of all compatible state paths is confined to the white (unshaded) portion of this graph.}}
\label{fig_zpaths}
\end{figure}

\begin{rem}
{Dobrushin's model of SEC (cf. Definition \ref{def_msec}) tracks the output string generated by each input symbol. In our model of the timing synchronization error channels, we track the input symbol that gave rise to each output symbol. We note that although the idea of modeling the ``drift'' between the transmitter and the receiver has been previously proposed, 
this has not been directly used to define a channel model. We show that many properties of this model can be exploited to arrive at bounds on the capacity of the channels with timing-synchronization errors.}
\end{rem}
{Fig. \ref{fig_zpaths} illustrates the way the channel $\mbf{P}_n$ operates.}
{The state process realization $z_{[n]}$ shown for $n = 10$ in Fig. \ref{fig_zpaths} is a compatible state path, since it corresponds to the state sequence $z_{[10]} = (0, -1, -1, 0, 1, 1, -1, -1, -2, -2)$ and a non-decreasing index process realization $\gamma_{[10]} = (1, 3, 4, 4, 4, 5, 8, 9, 11, 12)$. }
\textcolor{black}{This path produces the output $y_{[10]} = x_{\gamma_{[10]}}$ from the input $x_{[12]}$.}

{Equation \eqref{eq_eqchntp} says that the transition probability of the channel $\mathbf{P}_n$ is the sum over all} \textcolor{black}{compatible state paths satisfying the boundary conditions $\mathfrak{B}_0$} {producing output $\overline{y}$ from input $x_{[n]}$ of the probability of the state path times the probability of deleting the $(n - \gamma_{|\overline{y}|})$ trailing symbols of $x_{[n]}$. The following result, whose proof is deferred to Appendix \ref{app_chnequiv}, establishes the relation between the Dobrushin DRC $\mathbf{Q}_n$ and the {timing synchronization error channel} $\mathbf{P}_n$.\vspace{2mm}}

\begin{prop}[Channel Equivalence] \label{prop_chnequiv}
For each $n \in \mathbb{N}$, the channels $\mathbf{Q}_n$ and $\mathbf{P}_n$ are {identical}.\hfill$\blacksquare$\vspace{2mm}
\end{prop}
{Henceforth, we will use the terms timing synchronization error channels and DRCs interchangeably.} \textcolor{black}{We note the following useful results and omit their proofs since they follow as a consequence of Proposition \ref{prop_chnequiv}.}
\begin{cor}
\label{cor_idxset}
{For the channel $\mathbf{P}_n, n \in \mathbb{N}$, we have}
\begin{enumerate}[(i)]
\item {The} {ratio of the output length to the input length $\frac{N_n}{n} \rightarrow \frac{1 - \p{d}}{1 - \p{r}}$} {almost surely} {as $n \rightarrow \infty$.} \label{cor_lnprop3}
\item {By definition, $\{\Gamma_{[N_n]}\} \subseteq [n]$. We have $\msf{P}(i \notin \{\Gamma_{[N_n]}\}) = \p{d}$ for $i \in [n]$.} \label{cor_idxprob}
\item {The cardinality of the index set, $|\{\Gamma_{[N_n]}\}|$, is the number of symbols among $n$ input symbols that were not deleted at the output. We have }${\msf{E}\Big[|\{\Gamma_{[N_n]}\}|\Big]} {=} {n(1 - \p{d})}${. Note that $|\{\Gamma_{[N_n]}\}| \neq N_n$ in general when $\p{r} > 0$.} \hfill$\blacksquare$ \vspace{2mm}
\end{enumerate}
\end{cor}

\begin{cor}[Dobrushin's results for $\{\mathbf{P}_n\}_{n = 1}^{\infty}$] \label{cor_eqchn}
\vspace{2mm}For input $X_{[n]}$ and output ${Y_{[N_n]}}$ of the channel $\mbf{P}_n$, the quantity
\begin{align}
C &= \lim_{n \rightarrow \infty} \sup_{\msf{P}(X_{[n]})} \frac{1}{n} I(X_{[n]}; Y_{[N_n]} \textcolor{black}{\mid \mathfrak{B}_0}) \notag \\
&= \sup_{\mcl{X}_\mcl{M}} \lim_{n \rightarrow \infty} \frac{1}{n} I(X_{[n]}; Y_{[N_n]} \textcolor{black}{\mid \mathfrak{B}_0}), \notag
\end{align}
where $\mcl{X}_\mcl{M}$ represents stationary, ergodic, Markov processes over $\mbb{X}$, exists and is equal to the capacity of the sequence of channels $\{\mathbf{P}_n\}_{n = 1}^{\infty}$.\hfill$\blacksquare$\vspace{2mm}
\end{cor}
We will henceforth restrict our attention to this class of input processes. 
%

\begin{rem} \label{rem_gen}
\textcolor{black}{Variants of the channel model introduced here can be used to represent wider classes of channels. By considering the channel state process $\mathcal{Z}$ to be a higher-order Markov process, synchronization channels with memory such as the \emph{segmented} deletion channels \cite{liu_10_tit_segid, wan_11_all_segdel} can be modeled. Note however, that higher-order Markov processes can also be used to model some memoryless synchronization error channels, such as the \emph{elementary sticky channel} \cite{mit_08_tit_sticky} and Gallager's insertion-deletion channel \cite{gal_61_llr_seqdecsec}. \emph{Bounded} memoryless synchronization channels \cite{rah_11_arx_capid} can also be modeled via higher-order Markov $\mathcal{Z}$ processes. Some results on the achievable rates of such channels and cascades of such channels with subsitution error channels and inter-symbol interference (ISI) channels can be found in \cite{ven_11_arx_delins} and \cite{fer_11_tcom_insdelsub, hu_10_tcom_insdel}, respectively. Another set of channels that can also be modeled are the \emph{jitter} \cite{bag_93_the_jitter, arn_00_isit_jitter} and \emph{bit-shift} channels \cite{sha_91_tit_bitshift} as well as the \emph{write-channel} or \emph{grain-error} channels studied in the context of magnetic recording \cite{iye_11_tmag_wcm, maz_11_tit_1dgmr}. These latter channels can actually be seen as variants of the finite-state channel approximations we will introduce to the timing-synchronization channels in Section \ref{sec_drc}. \emph{Permuting} \cite{ben_75_the_perm}, \emph{Trapdoor} \cite{ash_65_bok_infoth} channels and \emph{Molecular Communication} channels \cite{cui_11_cwit_dsc} can also be thought of as variants of the timing synchronization error channel model defined here. Although each of the aforementioned cases are interesting in their own right, 
further consideration of these channels is beyond the scope of our paper. We emphasize that 
alternative channel state-based models might be of use in a wide context, but will stick to the DRC and its special cases henceforth.}
\end{rem}

\subsection{Bounds on the Capacity of the DRC} \label{ssec_ddcbounds}
The formulation of the DRC in Section \ref{ssec_chnmod} 
allows us to immediately establish the following 
\textcolor{black}{simple bounds on the capacity of the DRC for $(\p{d}, \p{r}) \in [0, 1)^2$:}
\begin{equation} \label{eq_simpbound}
(1 - \p{d})\Big(1 - \frac{h_2(\p{r})}{1 - \p{r}}\Big) - h_2(\p{d}) \leq C \leq 1 - \p{d}.
\end{equation}
{Here}
\begin{equation}
{h_2(x) \triangleq \begin{cases}-x\log_2x - (1 - x)\log_2(1 - x), &x \in (0, 1)\\ 0, &x \in \{0, 1\}\end{cases},}
\end{equation}
{is the \emph{binary entropy function} \cite{cov_06_bok_infoth}.}
{In fact, the lower bound above is actually a lower bound on the rate achieved by i.u.d. inputs, referred to as the \emph{symmetric information rate} (SIR) and denoted $C^{iud}$.} Three special cases of the DRC are of particular interest: the binary deletion channel (BDC) with $\p{d} = p, \p{r} = 0$; the \emph{symmetric} deletion-replication channel (SDRC) with $\p{d} = \p{r} = p$; and the binary replication channel (BRC) with $\p{d} = 0, \p{r} = p$. Specializing the bounds in \eqref{eq_simpbound} 
to these cases gives us the following.
\begin{align}
1 - p - h_2(p) &\leq {C_{\mathrm{BDC}}^{iud}} \leq C_{\mrm{BDC}} \leq 1 - p, \notag \\
1 - p - 2h_2(p) &\leq {C_{\mathrm{SDRC}}^{iud}} \leq C_{\mrm{SDRC}} \leq 1 - p, \label{eq_capbndsspl}\\
1 - \frac{h_2(p)}{1 - p} &\leq {C_{\mathrm{BRC}}^{iud}} \leq C_{\mrm{BRC}} \leq 1. \notag 
\end{align}
%
Although these bounds 
have simple closed-form expressions with well known information theoretic functions, they are loose compared to the best known (analytical or numerical) bounds for the capacity of these channels. {We note here that bounds similar to the above are shown to be tight for channels with large alphabets in \cite{mer_12_tit_nsc}.} We can, however, improve these bounds. \textcolor{black}{Using Corollary \ref{cor_idxset} \eqref{cor_idxprob}, it can be shown that}
\begin{align}
I(X_{[n]}; Y_{[N_n]} \textcolor{black}{\mid \mathfrak{B}_0}) 
&\geq (1 - \p{d})H(X_{[n]}) + I(Y_{[N_n]}; Z_{[N_n]} \textcolor{black}{\mid \mathfrak{B}_0}) \notag \\
&\hspace{5mm} - H(Z_{[N_n]} \textcolor{black}{\mid \mathfrak{B}_0}) + H(Z_{[N_n]} \mid X_{[n]}, Y_{[N_n]} \textcolor{black}{, \mathfrak{B}_0}). \label{eq_ixy}
\end{align}
\textcolor{black}{Further from the transition probabilities in \eqref{eq_ztp} and Corollary \ref{cor_idxset} \eqref{cor_lnprop3}, it is easy to see that}
\[
\textcolor{black}{H(Z_{[N_n]} \mid \mathfrak{B}_0) \doteq n\Big(\frac{1 - \p{d}}{1 - \p{r}}h_2(\p{r}) - h_2(\p{d})\Big).}
\]
\textcolor{black}{Consequently, we have}
\begin{align}
C \geq \sup_{\mcl{X}} \Big((1 - \p{d})&\mcl{H}(\mcl{X}) + \hat{\mcl{H}}(\mcl{Z} \mid \mcl{X}, \mcl{Y}, \textcolor{black}{\mathfrak{B}_0})\Big) 
- \frac{1 - \p{d}}{1 - \p{r}}h_2(\p{r}) - h_2(\p{d}), \label{eq_caplowbd}
\end{align}
where we have written the entropy rate of the input process $\mcl{X}$ as $\mcl{H}(\mcl{X})$ and defined
\begin{equation}
\hat{\mcl{H}}(\mcl{Z} \mid \mcl{X}, \mcl{Y}, \textcolor{black}{\mathfrak{B}_0}) \triangleq \lim_{n \rightarrow \infty} \frac{1}{n}H(Z_{[N_n]} \mid X_{[n]}, Y_{[N_n]}, \textcolor{black}{\mathfrak{B}_0}).
\end{equation}
%
%
We have
\begin{align}
\frac{1}{n} H(Z_{[N_n]} \mid X_{[n]}, Y_{[N_n]}, \mathfrak{B}_0) &= \frac{1}{n} H(Z_{[N_n]} \mid X_{[n]}, Y_{[N_n]}, N_n, \mathfrak{B}_0)\\
&= \frac{1}{n} \msf{E}\Big(\sum_{i = 1}^{N_n} H({Z_{i} \mid Z_{i - 1}, X_{[\Gamma_{i - 1} : n]}, Y_{[i : N_n]}, N_n, \mathfrak{B}_0})\Big). \label{eq_hnterm}
\end{align}
\textcolor{black}{For each $i \in [N_n]$, $Z_i$ is lower bounded by $i - n$ with probability $1$. In the limit $n \rightarrow \infty$, however, this is a trivial lower bound. For the term in the sum with $i = 1$ in Equation \eqref{eq_hnterm}, we have the additional information that $Z_1 \leq Z_0 = 0$, by the boundary condition $\mathfrak{B}_0$ in \eqref{eq_boundary_condn}. However, for $i \geq 2$, $\mathfrak{B}_0$ does not limit the state space of $Z_i$. From the fact that the transition probabilities \eqref{eq_ztp} of the $\mcl{Z}$ process are time-invariant, as $n \rightarrow \infty$, confining expectations to the set (of probability $1$) where $N_n \rightarrow \infty$,}
\begin{align}
H(Z_{i} &\mid Z_{i - 1}, X_{[\Gamma_{i - 1} : n]}, Y_{[i : N_n]}, N_n, \mathfrak{B}_0)  \notag \\
&\qquad\rightarrow H(Z_1 \mid Z_0 = 0, {X_0,} X_{\mbb{N}}, Y_{\mbb{N}}) \triangleq H(Z_1 \mid Z_0 = 0, X_0, \mcl{X}, \mcl{Y}). \notag
\end{align}
\textcolor{black}{We re-emphasize that the entropy term above is \emph{without} the conditioning on the boundary condition $N_0 = 0$. Here, we assume that $X_0$ is distributed according to the stationary marginal distribution of the $\mathcal{X}$ process.} Since $\frac{\msf{E}(N_n)}{n} \rightarrow \frac{1 - \p{d}}{1 - \p{r}}$, optimizing over input processes $\mcl{X}$ gives us
\begin{align}
C &\geq \sup_{\mcl{X}} \Big(\mcl{H}(\mcl{X}) + \frac{H(Z_1 \mid {Z_0 = 0, X_0,} \mcl{X}, \mcl{Y})}{1 - \p{r}}\Big)(1 - \p{d}) \label{eq_capbdclb}
-\frac{1 - \p{d}}{1 - \p{r}}h_2(\p{r}) - h_2(\p{d}). 
\end{align}
%
%
It is not easy to evaluate the above bound. 
However, we can further lower bound the capacity by introducing some conditioning.
\begin{lem} \label{lem_monli} \vspace*{2mm}
The sequence of lower bounds $\{D_i^{\mcl{X}}\}_{i = 1}^{\infty}$, where
\begin{align}
D_i^\mcl{X} &\triangleq \Big(\mcl{H}(\mcl{X}) + \frac{H(Z_1 \mid Z_i, {Z_0 = 0, X_0, } \mcl{X}, \mcl{Y})}{1 - \p{r}}\Big)(1 - \p{d}) \notag \\
&\qquad\qquad- \frac{1 - \p{d}}{1 - \p{r}}h_2(\p{r}) - h_2(\p{d}),{\ }i \in \mbb{N} \notag
\end{align}
is non-decreasing.
\end{lem}
\begin{IEEEproof}
Since we have introduced extra conditioning, the $D_i^\mathcal{X}$s are lower bounds. We have
\begin{align}
H(Z_1 &\mid Z_{i + 1}) = H(Z_1, Z_i \mid Z_{i + 1}) - H(Z_i \mid Z_1, Z_{i + 1}) \\
&
{=} H(Z_i \mid Z_{i + 1}) + H(Z_1 \mid Z_i) - H(Z_i \mid Z_1, Z_{i + 1}) \label{eq_hzmar} \\
&\geq H(Z_1 \mid Z_i) 
\end{align}
where 
\eqref{eq_hzmar} follows from the Markovity of the $\mcl{Z}$ process. Since conditioning on $(Z_0 = 0, X_0, \mcl{X}, \mcl{Y})$ preserves the above chain of inequalities, 
$\{D_i^{\mcl{X}}\}_{i = 1}^\infty$ is non-decreasing.
%
\vspace*{2mm}
\end{IEEEproof}
{Observing $Z_i$ essentially establishes synchronization at the time instant $i$ by spelling out the index of the input which is produced as the output at that instant. The idea that ``the longer we wait before establishing the synchronization, the harder it is to predict the input index of the first output'' is intuitive. This is formalized above by showing that the conditional entropy of $Z_1$ given $(Z_0 = 0, X_0, \mathcal{X}, \mathcal{Y})$ increases as the observation $Z_i$ is delayed. This fact is \textcolor{black}{used in} 
obtaining a sequence of lower bounds on the capacity of the DRC that are non-decreasing in $i$.} {Note that this approach is similar to the one in \cite{kir_10_tit_ddccap} wherein the authors try to estimate the entropy of the input sequence length producing an output run. However, the difference here is that by conditioning on $Z_i$, we are not imposing the constraint that the synchronization is established at the end of output runs, and that we are considering the entropy of the input symbol index corresponding to the first output symbol.}

%
\textcolor{black}{Starting from \eqref{eq_ixy}, without dropping the $I(Y_{[N_n]}; Z_{[N_n]} \mid \mathfrak{B}_0)$ term, and following similar steps as before, we can arrive at}
\begin{align}
C &\geq \sup_\mathcal{X}\Big( (1 - \p{d})\mathcal{H}(\mathcal{X}) - \hat{\mathcal{H}}(\mathcal{Z} \mid \mathcal{Y}, \mathfrak{B}_0) + \hat{\mathcal{H}}(\mathcal{Z} \mid \mathcal{X}, \mathcal{Y}, \mathfrak{B}_0)\Big) \\
&= (1 - \p{d})\Big[\sup_\mathcal{X}\Big(\mathcal{H}(\mathcal{X}) + \frac{H(Z_1 \mid Z_0 = 0, {X_0}, \mathcal{X}, \mathcal{Y}) - H(Z_1 \mid Z_0 = 0, {Y_0}, \mathcal{Y})}{1 - \p{r}}\Big)\Big]. \label{eq_capbrclb}
\end{align}
Following arguments similar to the ones used in Lemma \ref{lem_monli}, we can show the following.
\begin{lem} \label{lem_monri} \vspace*{2mm}
The sequence of lower bounds $\{R_i^\mathcal{X}\}_{i = 1}^{\infty}$, where
\begin{align}
R_i^\mathcal{X} \triangleq (1 - \p{d})\Big(\mathcal{H}(\mathcal{X}) &+ \frac{H(Z_1 \mid Z_i, Z_0 = 0, {X_0}, \mathcal{X}, \mathcal{Y}) - H(Z_1 \mid Z_0 = 0, {Y_0}, \mathcal{Y})}{1 - \p{r}}\Big) 
\end{align}
is non-decreasing.
\hfill$\blacksquare$
\end{lem}
{The bounds in Lemmas \ref{lem_monli} and \ref{lem_monri} have a couple of properties that render them amenable to analysis. Since the bounds are free of any maximal index time terms, the analysis in subsequent sections is largely simplified. Moreover, dealing with the conditional entropy of a single random variable as opposed to the conditional entropy rate of a process makes the derivations less cumbersome.}

\vspace*{2mm}The task of finding the rate-maximizing input distributions appears to be tough, with no theoretical insights\footnote{A new result on BDC with small deletion probability \cite{kan_13_tit_bdc} provides a partial answer to this question.} or efficient numerical algorithms. Often, to establish lower bounds on achievable rates, special classes of input processes are considered, and we will resort to a similar strategy here to obtain some expressions for the bounds we have so far developed. The following section will consider special cases of the DRC wherein there are either only deletions, i.e., the BDC, or only replications, i.e., the BRC. In a subsequent section, the symmetric DRC will be studied. 

\section{Channels with Deletions or Replications} \label{sec_dord}
For the case of the BDC or the BRC, evaluating some of the bounds developed in the previous section is somewhat easy, owing to the fact that the $\mcl{Z}$ process is monotonic in these two special cases, i.e., it is non-increasing or non-decreasing with increments of at most one, respectively. This monotonicity in $\mcl{Z}$ implies that the $\Gamma$ process is strictly increasing for the BDC and non-decreasing with increments of at most one for the BRC. This translates to the output being a subsequence of the input sequence for the BDC and vice versa for the BRC.

\subsection{Information Rates for the BDC} \label{ssec_bdc}
In this subsection we estimate the information rates possible over the BDC, i.e., $\p{d} = p, \p{r} = 0$, when the input process is either i.u.d. or when it is a first-order Markov process. 

For the BDC with i.u.d. inputs, we can easily show that $\mcl{Y}$ is also an i.u.d. sequence. 
\textcolor{black}{Combining with the fact that the $\varGamma$-process is strictly increasing for the BDC, we have $I(Y_{[N_n]}; Z_{[N_n]} \mid \mathfrak{B}_0) = 0$ so that the lower bound in \eqref{eq_capbdclb} is actually the SIR.}
We are hence interested in evaluating $D_i^\mrm{iud}$ as defined in Lemma \ref{lem_monli}, where the superscript ``iud'' stands for independent, uniformly distributed inputs. In particular, we have the SIR 
\begin{equation} \label{eq_sirbdc}
C^\mathrm{iud}_\mathrm{BDC} = \lim_{i \rightarrow \infty} D_i^\mathrm{iud} = \sup_{i \geq 1} D_i^\mathrm{iud}.
\end{equation}

We start with some definitions and notation.
\begin{defn}[Subsequence weights] \vspace*{2mm}
We call a vector $x_A$ a \emph{subsequence} of a vector $x_B$ if $A \subset B$ and the order of the elements in $A$ is the same as the order in which those elements appear in $B$. For ease of notation, we will write $w_{y_{[i]}}(x_{[j]})$ to denote the number of subsequences of $x_{[j]} \in \mathbb{X}^j$ that are the same as $y_{[i]} \in \mathbb{X}^i$, which is referred to as the $y_{[i]}$-\emph{subsequence weight} of the vector $x_{[j]}$. We can write
\[
w_{y_{[i]}}(x_{[j]}) = \sum_{S \subset [j] : |S| = i} \mathds{1}_{\{x_S = y_{[i]}\}}
\]
where the elements of the set $S$ are arranged in ascending order. Clearly, $w_{y_{[i]}}(x_{[j]}) = 0$ for $i > j$. We define $w_{\lambda}(x_{[j]}) = 1{\ }\forall{\ }x_{[j]} \in \mbb{X}^j$ for $j \geq 0$ \textcolor{black}{corresponding to the subsequence defined by $S = \emptyset$}.\hfill$\square$
\end{defn}
\begin{defn}[Runs and run-lengths] \vspace*{2mm}
For a binary sequence, a \emph{run} is a maximal block of contiguous $0$s or $1$s. The \emph{run-length} of a run is the number of symbols in it. We denote by $r_1(x_{[j]})$ the length of the first run in the vector $x_{[j]}\ \in \mbb{X}^j, j \geq 1$. Clearly, $1 \leq r_1(x_{[j]}) \leq |x_{[j]}| = j$.\vspace*{2mm}\hfill$\square$
\end{defn}

We will denote by $\mbb{Z}_{\uparrow}^{i}$ and $\mbb{Z}_{\downarrow}^i$ the sets of non-decreasing and non-increasing vectors of length $i$, respectively, for $i \geq 1$.

\begin{figure*}[!b]
\hrule
\begin{align} \label{eq_him}
&\mathfrak{H}^{(i)}_m = \sum_{x_{[m + i - 1]} \in \mbb{X}^{m + i - 1}}\frac{1}{2^{m + i - 1}}\msf{H}(x_{[m + i - 1]}), \notag \\
\msf{H}(x_{[m + i - 1]}) &= \sum_{y_{[i - 1]} \in \mbb{Y}^{i - 1}}\frac{w_{y_{[i - 1]}}(x_{[m + i - 1]})}{{m + i - 1 \choose m}}\mathfrak{h}(x_{[m + i - 1]}, y_{[i - 1]}), \\
\mathfrak{h}(x_{[m + i -1]}, y_{[i - 1]}) = -\sum_{z = -m}^0&\mds{1}_{\{x_{1 - z} = y_1\}}\frac{w_{y_{[2 : i - 1]}}(x_{[2 - z : m + i - 1]})}{w_{y_{[i - 1]}}(x_{[m + i - 1]})}\log_2\Big(\mds{1}_{\{x_{1 - z} = y_1\}}\frac{w_{y_{[2 : i - 1]}}(x_{[2 - z : m + i - 1]})}{w_{y_{[i - 1]}}(x_{[m + i - 1]})}\Big). \notag
\end{align}
\end{figure*}

\begin{thm}[SIR for the BDC] \label{thm_sirbdc} \vspace*{2mm}
For the BDC,
\begin{align}
C^\mathrm{iud}_\mathrm{BDC} &= 1 - p - h_2(p) \notag \\
&\quad + (1 - p)\Big(\lim_{i \rightarrow \infty} \sum_{m \geq 0} \psi_{i, m}p^m(1 - p)^i\Big), \notag
\end{align}
where $\psi_{i, m} \triangleq {{m + i - 1} \choose m}\mathfrak{H}^{(i)}_m$, with $\mathfrak{H}^{(i)}_m = H(Z_1 \mid Z_0 = 0, Z_i = -m,{\ }{X_0,}{\ }\mcl{X}, \mcl{Y})$ as given in Equation \eqref{eq_him}.
\end{thm}
\begin{IEEEproof}
For the BDC, we have from Lemma \ref{lem_monli} that
\begin{equation}
D_i^\mathrm{iud} = 1 - p - h_2(p) + (1 - p)H(Z_1 \mid Z_0 = 0, Z_i,{\ }{X_0,}{\ }\mcl{X}, \mcl{Y}).
\end{equation}
From Equation \eqref{eq_sirbdc}, we need to show that
\begin{equation}
H(Z_1 \mid Z_0 = 0, Z_i,{\ }{X_0,}{\ }\mcl{X}, \mcl{Y}) = \sum_{m \geq 0} \psi_{i, m}p^m(1 - p)^i.
\end{equation}
We first note that
\begin{equation}
H(Z_1 \mid Z_0 = 0, Z_i,{\ }{X_0,}{\ }\mcl{X}, \mcl{Y}) {= H(Z_1 \mid Z_0 = 0, Z_i, \mathcal{X}, \mathcal{Y})} = H(Z_1 \mid Z_0 = 0, Z_i, X_{[i - Z_i - 1]}, Y_{[i - 1]}),
\end{equation}
{since for the BDC, none of the outputs $Y_{\mathbb{N}}$ can correspond to $X_0$.} Clearly, the above entropy term is zero for $i = 1$. For $i \geq 2$, given $Z_0 = 0, Z_i = -m, X_{[m + i - 1]} = x_{[m + i - 1]}$ and $Y_{[i - 1]} = y_{[i - 1]}$, it is easy to see that
\begin{align}
Z_1~\in~&\{z \in \{0, -1, \cdots, -m\} : x_{1 - z} = y_1, \notag \\
&\qquad w_{y_{[2 : i - 1]}}(x_{[2 - z : m + i - 1]}) > 0\}. 
\end{align}
That is, $Z_1 = z$ only if $x_{1 - z}$ and $y_1$ match, and the subsequent part of the output vector $y_{[2 : i - 1]}$ is a subsequence of the subsequent part of the input vector $x_{[2 - z : m + i - 1]}$. 
\textcolor{black}{It is easy to see that}
 when $w_{y_{[i - 1]}}(x_{[m + i - 1]}) > 0$,
\begin{align}
\msf{P}(Z_1 = z \mid &Z_0 = 0, Z_i = -m, X_{[m + i - 1]} = x_{[m + i - 1]}, Y_{[i - 1]} = y_{[i - 1]}) \notag \\
&= \frac{\mds{1}_{\{x_{1 - z} = y_1\}}w_{y_{[2 : i - 1]}}(x_{[2 - z : m + i - 1]})}{w_{y_{[i - 1]}}(x_{[m + i - 1]})}. 
\end{align}
Since, with i.u.d. inputs, $\msf{P}(X_{[m + i - 1]} = x_{[m + i - 1]} \mid Z_i = -m) = 2^{-(m + i - 1)}$ and
\begin{align}
\msf{P}(Y_{[i - 1]} &= y_{[i - 1]} \mid X_{[m + i - 1]} = x_{[m + i - 1]}, Z_i = -m) \notag \\
&= \frac{w_{y_{[i - 1]}}(x_{[m + i - 1]})}{{m + i - 1 \choose m}}, 
\end{align}
we have that $H(Z_1 \mid Z_i = -m, \mcl{X}, \mcl{Y}) = \mathfrak{H}^{(i)}_m$ as in Equation \eqref{eq_him}. By noting that
\begin{equation}
\msf{P}(Z_i = -m \mid Z_0 = 0) = {{m + i - 1} \choose m}p^m(1 - p)^i
\end{equation}
from Equation \eqref{eq_ztp} (with $\p{d} = p, \p{r} = 0, \p{t} = 1 - p$), we have the desired result.\vspace*{2mm}
\end{IEEEproof}
\textcolor{black}{The subsequence weights in the expression for $\mathsf{H}$ can be easily evaluated through recursive counting techniques (See \cite{mit_09_prs_delchn} for a brief summary). The difficulty in evaluating $\mathfrak{H}^{(i)}_m$ is the exponential number of terms in the summation, which is computationally prohibitive for $i + m \geq 16$. However, }
we can evaluate it in two specific cases: for every $m$ when $i = 2$ (when all but a single bit are deleted) and for all $i$ when $m = 1$ (when only a single bit is deleted). We examine these two cases in detail in Appendix \ref{ssec_i2}, \ref{ssec_m1} and state the results here.

\begin{figure}[!ht]
\centering
\twofigs{\include{bdcbounds}}{\includegraphics{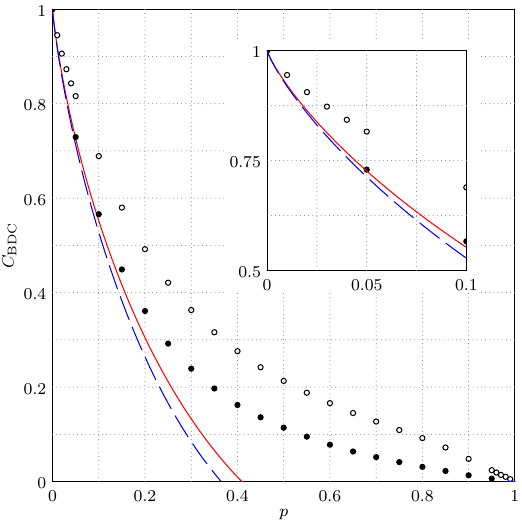}}
\caption{Bounds on the capacity for the BDC in bits per channel use as a function of the deletion probability $p$. $D_2^{\mrm{iud}}$ (cf. Corollary \ref{cor_l2}
) is shown as the long-dashed blue line and $C^\mathrm{iud}$ (or equivalently $\mathfrak{D}_1^{\mrm{iud}}$) with the $O(p^2)$ term dropped as the solid red line (cf. Equation \eqref{eq_ciudsdp}). \textcolor{black}{The best known numerical lower \cite{kir_10_tit_ddccap, mer_12_tit_nsc, mit_06_tit_lbcapdc} and upper bounds \cite{fer_10_tit_bdccap, dal_11_isit_capdelub, rah_13_all_bdcubp1} are shown as black and white circles respectively.} 
The inset shows the bounds for small $p$ values where the red solid curve is known to be tight from \cite{kan_10_isit_delchn}.}
\label{fig_bdc}
\end{figure}

\begin{cor}[Lower bound for $C^\mathrm{iud}_\mathrm{BDC}$] \label{cor_l2} \vspace*{2mm}
For the BDC,
\begin{align}
\hspace*{6mm}C^\mathrm{iud}_\mathrm{BDC} \geq D_2^\mathrm{iud} &\geq \frac{4(1 - p)^3}{(2 - p)^2} - h_2(p) \notag \\
&\quad + (1 - p)^3\Big(\sum_{m \geq 2}mp^{m - 1}\log_2m\Big). \tag*{$\blacksquare$}
\end{align}
\end{cor}
\begin{cor}[Small deletion probability SIR] \label{cor_sdpsir} \vspace*{2mm}
For the BDC,
\[
C^\mathrm{iud}_\mathrm{BDC} = 1 + p\log_2p - \mathtt{d}p + O(p^2)
\]
where $\mathtt{d} \approx 1.154163765$.\hfill$\blacksquare$
\end{cor}
%
Fig. \ref{fig_bdc} plots the bounds for $C_\mathrm{BDC}$.

\subsection{Information Rates for the BRC} \label{ssec_bdupc}
In this subsection, we will consider information rates for the BRC, i.e., $\p{d} = 0, \p{r} = p$. As in the previous subsection, we will consider i.u.d. and symmetric first-order Markov inputs. 
For the BRC, the $\mcl{Z}$ process is non-decreasing. Moreover, when it increases, the increment is at most $1$ at each time instant. This simplifies the evaluation of information rates and we will, in fact, be able to write exact expressions for the Markov-$1$ rates, as will be shown shortly. 

\begin{figure*}[!b]
\hrule
\begin{align}
C^{\mathcal{M}1}_\mathrm{BRC} = \max_{\alpha}\Big[h_2(\alpha) + \alpha\sum_{l \geq 1}\Big((1 - \alpha)\frac{1 - p}{p}\Big)^l\Big(\sum_{k \geq l}{k \choose l}p^kh_2(\frac{l}{k})\Big) - \frac{p + (1 - \alpha)(1 - p)}{1 - p}h_2\Big(\frac{p}{p + (1 - \alpha)(1 - p)}\Big) \Big]. \label{eq_m1rbrc}
\end{align}
\end{figure*}

\begin{thm}[Markov-$1$ Rates for the BRC] \label{thm_m1rbrc}\vspace*{2mm}
For the BRC, the Markov-$1$ rate is given as in Equation \eqref{eq_m1rbrc}.
\end{thm}
\begin{IEEEproof}
First we note that since $Z_1 \in \{0, 1\}$ \textcolor{black}{given $Z_0 = 0$}, we have $H(Z_1 \mid Z_0 = 0, {X_0}, \mcl{X}, \mcl{Y}) = \msf{E}[h_2(\msf{P}(Z_1~=~0 \mid Z_0 = 0, {x_0},~x_{\mbb{N}},~y_{\mbb{N}}))]$ and $H(Z_1\mid Z_0 = 0, {y_0},~\mcl{Y}) = \msf{E}[h_2(\msf{P}(Z_1~=~0\mid Z_0 = 0, {y_0},~y_{\mbb{N}}))]$. Further, we have for the BRC that whenever $Y_i \neq Y_{i - 1}$, we must have $Z_i = Z_{i - 1}$ or equivalently $\Gamma_i = \Gamma_{i - 1} + 1$. This means that $Z_1$ is independent of subsequent runs of $\mcl{Y}$ (and $\mcl{X}$) given the first run of $\mcl{Y}$ (and $\mcl{X}$) since we can achieve synchronization at the end of each run. Thus we can write the conditional probabilities $\msf{P}(Z_1~=~0~\mid Z_0 = 0, {X_0}, \mcl{X},~\mcl{Y})$ and $\msf{P}(Z_1~=~0\mid Z_0 = 0, {Y_0},\mcl{Y})$ in terms of the first runs of $\mcl{X}$ and $\mcl{Y}$, i.e., $\msf{P}(Z_1~=~0\mid Z_0 = 0, {x_0}, x_{\mbb{N}},~y_{\mbb{N}}) = \msf{P}(Z_1~=~0\mid Z_0 = 0, {x_0}, r_1(x_{\mbb{N}}),~r_1(y_{\mbb{N}}))$ and $\msf{P}(Z_1~=~0\mid Z_0 = 0, {y_0}, y_{\mbb{N}}) = \msf{P}(Z_1~=~0\mid Z_0 = 0, {y_0}, r_1(y_{\mbb{N}}))$. {Recall} that we assume $Z_0 = 0$ so that $Y_0 = X_0$. Thus, if $x_1 \neq x_0$, then $Z_1$ is $0$ or $1$ accordingly as $y_1$ is not equal or equal to $y_0$, respectively. This means that there is no uncertainty in $Z_1$ given the output sequence
. Therefore, in estimating the entropy of $Z_1$ given the output sequence, or {given} the output and the input sequences, we can confine our attention to those sequences $x_\mathbb{N}$ and $y_\mathbb{N}$ whose first runs are comprised of zeros with the assumption that $x_0 = y_0 = 0$, without loss of generality. We shall denote such runs as $r_1^0(\cdot)$. For a first-order Markov input process, we have, for $l \geq 0$
\begin{equation} \label{eq_mar1brcprob}
\msf{P}(r_1^0(x_{\mbb{N}}) = l {\mid x_0 = 0}) = (1 - \alpha)^l\alpha,
\end{equation}
and we can get from the definition of the BRC that
\begin{equation}
\msf{P}(r_1^0(y_{\mbb{N}}) = k \mid {x_0 = 0}, r_1^0(x_{\mbb{N}}) = l) = {k \choose l}(1 - p)^{l + 1}p^{k - l}
\end{equation}
for $k \geq l$. Consequently, {and since $Y_0 = X_0$}, we have
\begin{align}
\msf{P}(r_1^0(y_{\mbb{N}}) 
= \alpha(1 - p)\Big(p + (1 - \alpha)(1 - p)\Big)^k. 
\end{align}
Since $Z_1 = 0$ excludes the first bit in the received sequence from being a replication, we can easily obtain
\begin{equation}
\msf{P}(Z_1 = 0 \mid {x_0 = 0}, r_1^0(x_{\mbb{N}}) = l, r_1^0(y_{\mbb{N}}) = k) = \frac{l}{k}
\end{equation}
for $k \geq l + \mds{1}_{\{l = 0\}}$. For $k \geq 1$, we can show that
\begin{align}
\msf{P}(Z_1 = 0 
\mid {y_0 = 0}, r_1^0(y_{\mbb{N}}) = k) 
&= \frac{(1 - \alpha)(1 - p)}{p + (1 - \alpha)(1 - p)}. 
\end{align}
Therefore,
\begin{align}
H(Z_1 
\mid {X_0}, \mcl{X}, \mcl{Y}) 
&= \alpha(1 - p)\sum_{l \geq 1} \Big((1 - \alpha)\frac{1 - p}{p}\Big)^l \Big(\sum_{k \geq l}{k \choose l}p^kh_2(\frac{l}{k})\Big) 
\end{align}
and
\begin{align}
H(Z_1 \mid {Y_0}, \mcl{Y}) 
&= (p + (1 - \alpha)(1 - p))h_2\Big(\frac{p}{p + (1 - \alpha)(1 - p)}\Big). 
\end{align}
Substituting these in \eqref{eq_capbrclb} 
specialized to the BRC and first-order Markov inputs, we have the desired result.
\end{IEEEproof}
\vspace*{2mm}The following results are shown in Appendix \ref{app_brc}.

\begin{cor}[Lower bound for $C^{\mathcal{M}1}_\mathrm{BRC}$] \label{cor_brclb}\vspace*{2mm}
For the BRC,
\begin{align}
C^{\mathcal{M}1}_\mathrm{BRC} \geq R_2^{\mathcal{M}1} &= h_2\Big(\frac{1}{(1 - p)(4^p + 1)}\Big) 
+ \Big(\frac{2p}{1 - p}\Big)\Big(\frac{(1 - p)4^p - p}{4^p + 1}\Big) 
- \Big(\frac{1}{1 - p}\Big)\Big(\frac{4^p}{4^p + 1}\Big)h_2\Big(\frac{p(4^p + 1)}{4^p}\Big) \notag 
\end{align}
for $0 \leq p \leq p_* \approx 0.734675821$.\hfill$\blacksquare$\vspace*{2mm}
\end{cor}

\begin{figure}[!ht]
\centering
\twofigs{\include{stibounds}}{\includegraphics{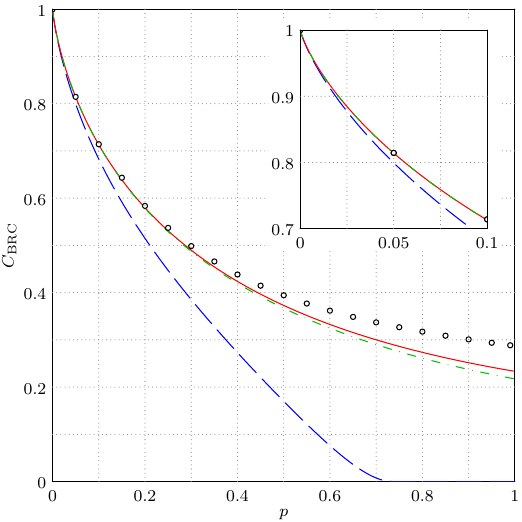}}
\caption{Lower bounds on the capacity for the BRC. The bound $R_2^{\mathcal{M}1}$ from Corollary \ref{cor_brclb} is shown as the long-dashed blue line and the Markov-$1$ rate in Equation \eqref{eq_m1rbrc} is shown as the solid red line. The SIR ($\alpha = \frac{1}{2}$ in Equation\eqref{eq_m1rbrc}) is the dash-dotted green line. 
The inset shows the bounds for small replication probabilities. {Note that the Markov-$1$ rate in solid red matches with the estimates in \cite{dri_07_tit_implbddc,mer_12_tit_nsc}. Shown as 
\textcolor{black}{dots are the 
numerically evaluated 
capacity 
from \cite{mer_12_tit_nsc}.}}}
\label{fig_brc}
\end{figure}

\begin{cor}[Small replication probability SIR] \label{cor_srpbrc} \vspace*{2mm}
For the BRC,
\[
C^\mathrm{iud}_\mathrm{BRC} = 1 + p\log_2p + \mathtt{r}p + O(p^2)
\]
where $\mathtt{r} \approx 0.845836235$.\hfill$\blacksquare$\vspace*{2mm}
\end{cor}
Fig. \ref{fig_brc} plots these bounds.
{Note that the Markov-$1$ rate evaluated here exactly agrees with the numerically evaluated rates 
in \cite{mer_12_tit_nsc}. 
Although sharper bounds on the capacity were provided in \cite{mer_12_tit_nsc}, we note that these bounds were all arrived at numerically
. Interestingly, although the expression derived here is itself different from the expression for the Markov-$1$ rate in \cite{dri_07_tit_implbddc}, the numerical values agree. Our result therefore represents an alternative 
proof technique for the rates achieved by first-order Markov processes over the BRC.}
\begin{rem}
{Note that the generalization of the results presented here to Markov processes of higher order is straightforward. The only changes to be made are in Equation \eqref{eq_mar1brcprob} that tracks the probability of input runs, and the entropy rate of the input Markov process. The advantage with first-order Markov processes is that the corresponding probability of output runs also turns out to have a simple closed-form expression, resulting in a simplified expression for the rate achieved.}
\end{rem}

%

\section{Channels with Deletions and Replications} \label{sec_drc}
Although the bounds in the previous section provide us some idea of the achievable information rates for the BDC and the BRC, they do not generalize in a straightforward manner for an SEC with both deletions and replications\footnote{It is possible to obtain, albeit with a lot more effort than in the cases of the BDC or the BRC, the lower bound $D_2^{\mathcal{X}}$ for a first-order Markov input process for a DRC. We shall omit this here.}. In order to obtain bounds when both deletions and replications are present, we take a different approach. 
{We attempt to approximate the achievable rates on the DRC by judiciously constructing a sequence of finite-state channels for which methods of optimizing input distributions are known. In the next two subsections, we highlight the general approach and discuss its advantages and limitations.}

\textcolor{black}{A straightforward way to obtain a finite-state channel (FSC) that approximates the timing synchronization error channels is to restrict the state space by clipping the $\mathcal{Z}$-process, i.e., by defining the channel state}
\begin{equation}
Z_i^{(m)} \triangleq \Big(\mds{1}_{\{|Z_i| \leq m\}}\Big)Z_i + \Big(\mds{1}_{\{|Z_i| > m\}}\Big)m\cdot \mrm{sgn}(Z_i), i \in \mathbb{Z}, \label{eq_zm}
\end{equation}
\textcolor{black}{where $\mrm{sgn}(\cdot)$ is the sign function. Writing $\Gamma_i^{(m)}$ and $N_n^{(m)}$ based on the so-obtained process $\mathcal{Z}^{(m)} \triangleq \{Z_i^{(m)}\}_{i\in\mathbb{Z}}$, it can be shown that the FSCs $\mathbf{P}_{n, m}^\dagger$ are \emph{totally ordered} in terms of their mutual information rates. That is, for a fixed $n \in \mathbb{N}$, we can show that for any $m \in \mbb{N}$,}
\begin{equation}
I(X_{[n]}; Y_{[N_n]}) \leq I(X_{[n]}; Y^{(m)}_{[N^{(m)}_n]}) \leq I(X_{[n]}; Y^{(m - 1)}_{[N^{(m - 1)}_n]}).
\end{equation}
\textcolor{black}{Despite this useful property, the channels $\{\mathbf{P}_{n, m}^\dagger\}_{m\geq 0}$ are not useful in evaluating bounds on the achievable rates of the timing synchronization error channel $\mathbf{P}_n$.}
{This is because $Z_n$, \textcolor{black}{which can be written as} 
the $n^\text{th}$-partial sum of i.i.d. random variables
, has a variance that grows with $n$. Consequently, the probability of $|Z_n| < m$ goes to zero for any $m \in \mathbb{Z}^+$, so that the $\mathcal{Z}^{(m)}$ process loses information of the $\mathcal{Z}$ process \textcolor{black}{unless $m = \Omega(n)$, i.e., the channel is no more a ``finite-state'' channel.}} 

\subsection{{Constructing Indecomposable FSCs}}
\label{ssec_appsc}
\textcolor{black}{We now }{construct FSCs for which the achievable information rates can be estimated. 
\textcolor{black}{The channel model for the FSC indexed by $m \in \mathbb{Z}^+$ is given by}
\begin{equation}
Y_i^{(m)} = X_{\Gamma_i^{(m)}} = X_{i - Z_i^{(m)}}, i \in \mathbb{Z},
\end{equation}
\textcolor{black}{where the channel state $Z_i^{(m)}$ is as defined in Equation \eqref{eq_zm}. We define} the measure $\mathsf{P}_{\langle m\rangle}$ \textcolor{black}{governing the channel behavior} 
such that the $\mathcal{Z}^{(m)}$ process is a finite, time-homogeneous, first-order Markov chain with transition probabilities
\begin{equation}
\mathsf{P}_{\langle m\rangle}(Z_i^{(m)} = k \mid Z_{i - 1}^{(m)} = j) = \mathsf{P}(Z_i^{(m)} = k\mid Z_{i - 1}^{(m)} = j)
\end{equation}
when $-m < j < m$, and
\begin{align}
\mathsf{P}_{\langle m\rangle}(Z_i^{(m)} = k \mid 
Z_{i - 1}^{(m)} = -m) 
&=
\begin{cases}
1 - \p{r}, &k = -m\\
\p{r}, &k = -m + 1\\
0, &\text{otherwise},
\end{cases}
\end{align}
and
\begin{align}
\mathsf{P}_{\langle m\rangle}
(Z_i^{(m)} = k \mid Z_{i - 1}^{(m)} = m) 
&=
\begin{cases}
1 - \p{d}(1 - \p{r}), &k = m\\
(1 - \p{d})(1 - \p{r})\p{d}^{m - k}, &-m < k < m\\
(1 - \p{r})\p{d}^{2m}, &k = -m\\
0, &\text{otherwise}.
\end{cases}
\end{align}
The transition probabilities $P_{n, m}^\star$ for the channel $\mathbf{P}_{n, m}^\star$ can now be defined as in Equation \eqref{eq_eqchntp}, but under the measure $\mathsf{P}_{\langle m\rangle}$. 

\begin{rem} \vspace*{2mm}
Note that the sequence of sigma-algebras $\{\mathscr{G}_m\}_{m \geq 0}$ where $\mathscr{G}_m \triangleq \sigma(\mathcal{Z}^{(m)})$ forms a filtration. The sequence of measures $\mathsf{P}_{\langle m\rangle}$ as defined above seem to be defined only on the corresponding sigma-algebras $\mathscr{G}_m$s for each $m \in \mathbb{Z}^+$. However, we can extend these measures to the sigma-algebra $\mathscr{B}$ as in Appendix \ref{app_extmeas}, and will henceforth consider $\mathsf{P}_{\langle m\rangle} : \mathscr{B} \mapsto [0, 1]$ for each $m \in \mathbb{Z}^+$.\vspace*{2mm}\hfill$\square$
\end{rem}
The lemma below shows that for a fixed $m \in \mathbb{Z}^+$, the FSC $\mathbf{P}_{n, m}^\star$ is an indecomposable FSC \cite[\S 4.6]{gal_68_bok_infoth}.
\begin{lem}[$\mathbf{P}_{n, m}^\star$ Indecomposable] \label{lem_indec}\vspace*{2mm}
The FSC $\mathbf{P}_{n, m}^\star$ is indecomposable for every $m \in \mathbb{Z}^+$ for $(\p{d}, \p{r}) \in (0, 1)^2$.
\end{lem}
\begin{IEEEproof}
Fix $m \in \mathbb{Z}^+$. We need to make a couple of modifications to put the channels $\{\mathbf{P}_{n, m}^\star\}_{n \geq 1}$ in the parlance of discrete FSCs. First, we set
\begin{equation}
\acute{Y}_i^{(m)} = Y_{i - m}^{(m)} = X_{i - m - Z_{i - m}^{(m)}} = X_{i - \acute{Z}_i^{(m)}} \textcolor{black}{, i \in \mathbb{Z}}
.
\end{equation}
Note that $\acute{Z}_i^{(m)} = m + Z_{i - m}^{(m)} \in [0 : 2m]$, and hence the channel producing $\acute{Y}_i
^{(m)}$ is ``causal''. Let the ``state'' $W_i^{(m)}$ of the channel $\mathbf{P}_{n, m}^\star$ at time $i \in \mathbb{Z}$ 
be defined as
\begin{equation}
W_i^{(m)} \triangleq (X_{[i - 2m : i - 1]}, \acute{Z}_i^{(m)}) \in \mathbb{X}^{2m} \times [0 : 2m].
\end{equation}
Note that we need to redefine the state of the channel in this case to keep the factorization
\begin{align}
\mathsf{P}_{\langle m\rangle}(&\acute{Y}_i^{(m)}, W_{i + 1}^{(m)} \mid X_i, W_i^{(m)})
= \mathsf{P}_{\langle m\rangle}(\acute{Y}_i^{(m)} \mid X_i, W_i^{(m)}
)\cdot\mathsf{P}_{\langle m\rangle}(W_{i + 1}^{(m)} \mid X_i, W_i^{(m)}
). 
\end{align}
%
Since $\acute{\mathcal{Z}}^{(m)}$ is a finitely delayed, finitely shifted version of $\mathcal{Z}^{(m)}$, and because $\mathcal{Z}^{(m)}$ is an \emph{irreducible, aperiodic} Markov chain \cite[Chapter 1]{law_06_bok_stocproc} under the measure $\mathsf{P}_{\langle m\rangle}$ as long as $(\p{d}, \p{r}) \in (0, 1)^2$, so is $\acute{\mathcal{Z}}^{(m)}$. In particular, we have that for every $j \geq 2m$,
\begin{equation}
\min_{z \in [0 : 2m]}\mathsf{P}_{\langle m\rangle}(\acute{Z}_{i + j}^{(m)} = z \mid \acute{Z}_i^{(m)} = z^\prime) > 0{\ }\forall{\ }z^\prime \in [0 : 2m].
\end{equation}
This implies that for \textcolor{black}{$i = 1, j = 2m$} and $\overline{x} \in \overline{\mathbb{X}}$, by choosing 
$w = (x_{[2m]}, z)$, for any $z \in [0 : 2m]$, we see that
\begin{equation}
\mathsf{P}_{\langle m\rangle}(W_{2m + 1}^{(m)} = w \mid \overline{X} = \overline{x}, W_1^{(m)} = w^\prime) > 0
\end{equation}
for every $w^\prime \in \mathbb{X}^{2m}\times [0 : 2m]$. From \cite[Theorem 4.6.3]{gal_68_bok_infoth}, we have the desired result.\vspace*{2mm} 
\end{IEEEproof}
\textcolor{black}{A consequence of the above result and the fact that the channel input alphabet $\mathbb{X}$ is finite is that the boundary condition imposed on the $\mathcal{W}^{(m)}$ process will not influence the capacity of the indecomposable FSC \cite[Theorem 4.6.4]{gal_68_bok_infoth}. }
\textcolor{black}{We will therefore assume the boundary condition $\mathfrak{B}_0^{(m)} \triangleq \{Z_0^{(m)} = 0, N_0^{(m)} = 0\}$. It is clear that the maximal index times for the $\mathcal{Z}^{(m)}$ process satisfy $|N_n^{(m)} - n| \leq m$ for $n \in \mathbb{N}$ under $\mathfrak{B}_0^{(m)}$. Note that by definition
, the measure $\mathsf{P}_{\langle m\rangle}$ differs from $\mathsf{P}$ only for state paths that reach beyond the states $\pm(m - 1)$. The following result follows immediately from \cite{fei_59_inc_capfsc}.}
\begin{cor}[Capacity of $\mathbf{P}_{n, m}^\star$] \label{cor_marcapfsc}
For the FSCs $\{\mathbf{P}_{n, m}^\star\}_{n \geq 1}$, the capacity $C^\star{(m)}$ can be written as
\begin{align}
C^\star{(m)} &= \sup_\mathcal{X} \lim_{n \rightarrow \infty}\frac{1}{n}I(X_{[n]}; Y_{[N_n^{(m)}]}^{(m)} \mid \mathfrak{B}_0^{(m)}) 
= \sup_\mathcal{X} \lim_{n \rightarrow \infty} I_{n, m}^\star \triangleq \sup_{\mathcal{X}} I_\mathcal{X}^\star(m) 
\end{align}
where the supremum is over all stationary, ergodic input sources $\mathcal{X}$.\vspace*{2mm}\hfill$\blacksquare$
\end{cor}
From Lemma \ref{lem_indec}, since $\{\mathbf{P}_{n, m}^\star\}_{n \geq 1}$ are indecomposable FSCs, we have from \cite{bla_58_anms_aepfsic} that
\begin{align}
-\frac{1}{n}\log_2\mathsf{P}_{\langle m\rangle}(X_{[n]}, Y_{[{N_0^{(m)} + 1} : N_n^{(m)}]}^{(m)}) &\rightarrow \lim_{n \rightarrow \infty}\frac{H(X_{[n]}, Y_{[{N_0^{(m)} + 1} : N_n^{(m)}]}^{(m)})}{n} 
= \hat{\mathcal{H}}(\mathcal{X}, \mathcal{Y}^{(m)}), \\
-\frac{1}{n}\log_2\mathsf{P}_{\langle m\rangle}(Y_{[{N_0^{(m)} + 1} : N_n^{(m)}]}^{(m)}) &\rightarrow \lim_{n \rightarrow \infty}\frac{H(Y_{[{N_0^{(m)} + 1} : N_n^{(m)}]}^{(m)})}{n} 
= \hat{\mathcal{H}}(\mathcal{Y}^{(m)}), 
\end{align}
as $n \rightarrow \infty$ almost surely, where the entropies are calculated with respect to the measure $\mathsf{P}_{\langle m\rangle}$. Therefore,
\begin{equation}
I_\mathcal{X}^\star(m) = \mathcal{H}(\mathcal{X}) + \hat{\mathcal{H}}(\mathcal{Y}^{(m)}) - \hat{\mathcal{H}}(\mathcal{X}, \mathcal{Y}^{(m)})
\end{equation}
can be estimated numerically using the forward passes of the BCJR algorithm \cite{bah_74_tit_bcjr} to estimate $\hat{\mathcal{H}}(\mathcal{X}, \mathcal{Y}^{(m)})$ and $\hat{\mathcal{H}}(\mathcal{Y}^{(m)})$, as in \cite{arn_06_tit_irchnmem, pfi_01_glb_achisi}. 
In Fig. \ref{fig_sdrc}, we plot the SIRs, $C_\mathrm{iud}^\star(m)$, for the indecomposable FSCs $\{\mathbf{P}_{n, m}^\star\}_{n \geq 1}$ obtained through numerical simulations for {$m = 2^k, k \in \{2, 3, 4, 5\}$ and $\p{d} = \p{r} = p \in [0, 1)$}.
\begin{figure}[!ht]
\centering
\twofigs{\include{sdrc}}{\includegraphics{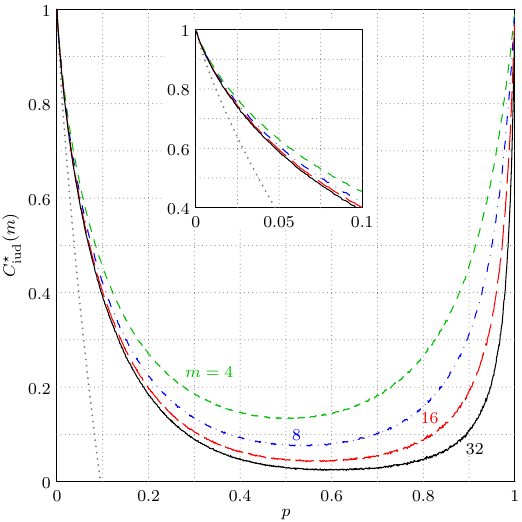}}
\caption{SIR estimates for the FSCs $\{\mathbf{P}_{n, m}^\star\}_{n \geq 1}$ with {$\p{d} = \p{r} = p \in [0, 1)$} for different $m$ values. The lower bound on the {SIR} of the SDRC from Equation \ref{eq_capbndsspl} 
is also shown as the gray-dotted line. {The inset shows the regime where we expect $C^\star_{\mathrm{iud}}(32)$ to be close to $C^\star_{\mathrm{iud}}(\infty)$.}}
\label{fig_sdrc}
\end{figure}
The value of $n$ used for the estimation was {$10^6$}. 

A couple of observations are worthwhile noting. First, the SIRs $\{C_\mathrm{iud}^\star(m)\}_{m \geq 0}$ are non-increasing. This hints at a total ordering of the FSCs $\{\mathbf{P}_{n, m}^\star\}_{m \geq 0}$ with respect to the information rates similar to \textcolor{black}{what can be shown for the channels $\{\mathbf{P}_{n, m}^\dagger\}_{m\geq 0}$}. 
Second, we see that for small values of $p$, the SIRs get bunched up as $m$ increases, i.e., the SIRs $C_\mathrm{iud}^\star(m)$ converge quickly, so that we have a good estimate of
\begin{equation}
C_\mathrm{iud}^\star(\infty) \triangleq \lim_{m \rightarrow \infty} C_\mathrm{iud}^\star(m)
\end{equation}
for $p$ close to $0$. {However, the estimates are loose for $p$ close to $1$. This is because for any $m \in \mathbb{N}$, as $p$ approaches $1$, the stationary distribution of the $\mathcal{Z}^{(m)}$ process under the measure $\mathsf{P}_{\langle m\rangle}$ converges to a distribution with all the mass on the state $m$, and consequently, the channel converges to a noiseless channel. On the other hand, the SDRC becomes more noisy as $p$ approaches $1$.}

\begin{prop} \label{prop_inmstar} \vspace*{2mm}
For $n \in \mathbb{N}$, we have
\begin{equation}
I_n^\star \triangleq \liminf_{m \rightarrow \infty} I_{n, m}^\star = I_n.
\end{equation}
Thus,
\begin{equation}
C = \sup_\mathcal{X} \inf_{n \geq 1} \liminf_{m \rightarrow \infty}I_{n, m}^\star.
\end{equation}
\end{prop}
\begin{IEEEproof}
For a fixed $n \in \mathbb{N}$, since we have that $\mathsf{P}_{\langle m\rangle}(X_{[n]}, Y_{[N_n^{(m)}]}^{(m)} \mid \mathfrak{B}_0^{(m)}) \rightarrow \mathsf{P}(X_{[n]}, Y_{[N_n]} \mid \mathfrak{B}_0)$ as $m \rightarrow \infty$ for every $\vartheta \in \mathbb{S}$, we also have the convergence 
in total variation
. Consequently, from \cite[Corollary 1$^\prime$]{dob_60_tproba_limexp}, we have the desired result. \vspace*{2mm}
\end{IEEEproof}

\begin{rem} \vspace*{2mm} \label{rem_conj2}
We conjecture that $I_{n, m}^\star \rightarrow I_n^\star = I_n$ as $m \rightarrow \infty$. From \cite[Corollary 1$^\prime$]{dob_60_tproba_limexp}, one needs to show uniform integrability of the \emph{information densities} $i(X_{[n]}, Y_{[{N_0^{(m)} + 1} : N_n^{(m)}]}^{(m)})$ for the conjecture to be true. Alternatively, if the sequence of channels $\{\mathbf{P}_{n, m}^\star\}_{m \geq 0}$ is totally ordered for every $n \in \mathbb{N}$ with respect to the mutual information rates, 
i.e., if $\{I_{n, m}^\star\}_{m \geq 0}$ is a non-increasing sequence for every $n \in \mathbb{N}$, then we know that
\begin{equation}
\lim_{m \rightarrow \infty} I_{n, m}^\star = I_n^\star,
\end{equation}
and from Proposition \ref{prop_inmstar}, $I_{n, m}^\star \downarrow I_n^\star$ follows.
 Unfortunately, we are not able to show this monotonicity in the sequence $\{I_{n, m}^\star\}_{m \geq 0}$. 
However, Fig. \ref{fig_sdrc} provides sufficient empirical evidence for this monotonicity conjecture.\vspace*{2mm}\hfill$\square$
\end{rem}

\subsection{Approximating Channels for the SDRC}
In this subsection, we consider the SDRC, i.e., the case when $\p{d} = \p{r} = p \in [0, 1)$. This channel is of interest since in practice, systems prone to mis-synchronization are usually not biased to produce more deletions or replications. For the case of the SDRC, we can \textcolor{black}{strengthen the result from the previous subsection. We leave the proof to Appendix \ref{app_sdrclim}.}
%

\begin{prop}[Approximating SDRC] \label{prop_sdrcap} \vspace*{2mm}
For the SDRC,
\[
I_{\mathcal{X}} = \lim_{n \rightarrow \infty} I_n = \liminf_{n \rightarrow \infty} I_{n, m(n)}^\star
\]
where $m(n) = \omega(\sqrt{n})$, for stationary, ergodic input process $\mathcal{X}$.\vspace*{2mm}\hfill$\blacksquare$
\end{prop}

The channels $\{\mathbf{P}_{n, m}^\star\}_{m \geq 0}$ give us a way to approach the problems of optimizing input distributions as well as designing coding schemes for the SDRC. We can optimize the inputs of $\mathbf{P}_{n, m}^\star$, starting with small values of $m$, under some input assumptions, e.g., for fixed-order Markov inputs \cite{kav_01_glb_marcap, von_08_tit_gbaa}. Note that the numerical estimation of $I_{n, m}^\star$ is possible (as described in the previous subsection) only when $m < n$, since setting the channels as indecomposable FSCs (cf. Lemma \ref{lem_indec}) is possible only in this case. Moreover, for a good estimate of the information rate, we will require $m \ll n$. For the SDRC, Proposition \ref{prop_sdrcap} allows us to consider some $\mathbf{P}_{n, m(n)}^\star$, where $m(n)$ is both $\omega(\sqrt{n})$ as well as $o(n)$, for which a good estimate of the information rate $I_{n, m(n)}^\star$ can be obtained. Note that due to the lack of a result analogous to Lemma \ref{lem_prub} in the case of a general DRC for $m < n$, generalizing these arguments when $\p{d} \neq \p{r}$ is not completely justified.

Starting with some small values of $m$, we expect that the information rates and optimal distributions quickly converge (in $m$), giving us a way to characterize optimal inputs for the SDRC $\mathbf{P}_n$. For small values of $p$, as in Fig. \ref{fig_sdrc}, the information rates for the SDRC can be characterized numerically for moderate values of $m$ (much smaller than $\omega(\sqrt{n})$ guaranteed by Lemma \ref{lem_prub}). For optimizing the input distribution for an approximation $\mathbf{P}_{n, m}^\star$, we can start with optimizing inputs that are $\mu^\text{th}$-order Markov processes, for $\mu \geq 1$. As was observed\footnote{Although the validity of the bounds in \cite{vve_68_ppi_capcomp} is unclear (See, e.g., \cite{dri_07_tit_implbddc}), the rapid convergence of information rates as a function of the order $\mu$ of the input Markov process is expected to be true.} in \cite{vve_68_ppi_capcomp}, the convergence of optimal information rates as a function of the order $\mu$ of the input Markov process is expected to be rapid. The authors in \cite{vve_68_ppi_capcomp} hypothesized that this convergence was exponential in $\mu$. Similar ``diminishing returns'' on increasing $\mu$ has also been observed by others \cite{kav_01_glb_marcap, von_08_tit_gbaa}. We think that a similar rapid convergence of $I_{n, m}^\star(\mathcal{X}_{\mathcal{M}\mu}^*)$ to $C_n(\mathcal{X}_{\mathcal{M}\mu}^*)$ also holds for $m$, where $I_{n, m}^\star(\mathcal{X}_{\mathcal{M}\mu}^*)$ is the optimal information rate achieved by a $\mu^\text{th}$-order Markov input process on the FSC $\mathbf{P}_{n, m}^\star$ and $C_n(\mathcal{X}_{\mathcal{M}\mu}^*)$ is the optimal information rate achieved by a $\mu^\text{th}$-order Markov input process on the SDRC $\mathbf{P}_n$. We use the \emph{generalized Blahut-Arimoto algorithm} presented in \cite{von_08_tit_gbaa} to evaluate $I_{n, m}^\star(\mathcal{X}_{\mathcal{M}\mu}^*)$ for some small values of $m$ and $\mu$. Fig. \ref{fig_sdrccap} plots these estimates, which illustrates the aformentioned observations.
\begin{figure}[!ht]
\centering
\twofigs{\include{sdrcplot}}{\includegraphics{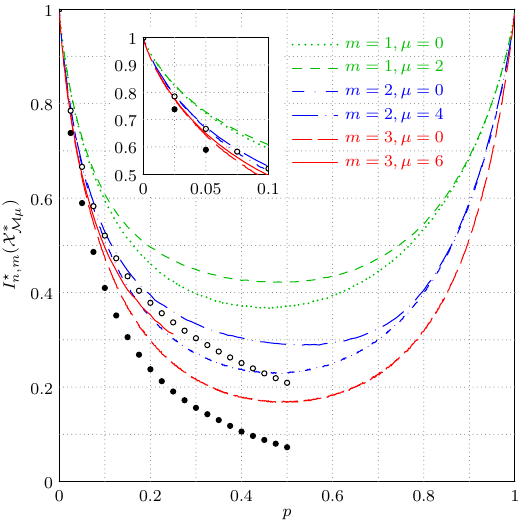}}
\caption{Numerical estimates of $I_{n, m}^\star(\mathcal{X}_{\mathcal{M}\mu}^*)$ for $m = 1, 2$ and $3$, and $\mu = 2m$ (solid lines). For the case where $m = 3, \mu = 6$, only the estimates for small deletion-replication probabilities have been evaluated. We chose $n = 10^6$ for $m = 1, 2$ and $n = 10^5$ for $m = 3$. The smaller value of $n$ in the case of $m = 3$ was chosen for computational convenience. Also shown for comparison are the corresponding estimates of the SIR ($\mu = 0$) from Fig. \ref{fig_sdrc}. {The best-known numerical lower and upper bounds on the capacity of the SDRC from \cite{mer_12_tit_nsc} are shown as black and white circles, respectively.}}
\label{fig_sdrccap}
\end{figure}
Note that it is clear from the plots in Fig. \ref{fig_sdrccap} that Bernoulli equiprobable inputs achieve rates (SIR) comparable to higher-order Markov inputs for small proababilities of deletion and replication. 
It is also evident from the figure that for small values of the deletion-replication probabilities, the information rates seem to converge as $m$ increases even for small values of $m$ ($m = 3$). This suggests that Fig. \ref{fig_sdrccap} plots good estimates of $I_{n, m(n)}^\star$ for such channels, and since the values of $n$ chosen are large, that they indeed represent good estimates of the Markov capacity of the SDRC. {This is substantiated by the closeness of the estimates to the bounds from \cite{mer_12_tit_nsc} in this regime. However, as was the case for the SIR estimates, the convergence is slower as $p$ increases.} 
\textcolor{black}{Note that restricting the state space of the $\mathcal{Z}$ process is equivalent to bounding the drift between the channel input and output--an approach which was used in \cite{mer_12_tit_nsc} to arrive at the bounds included in Fig. \ref{fig_sdrccap}. Our treatment here establishes the theoretical background for the convergence and the rate of convergence of information rates for channels with growing drift between input and output.}

Apart from the above advantage of facilitating numerical estimation of information rates, the approximating channels $\mathbf{P}_{n, m}^\star$ have another important advantage. This is that since they have immediate factor-graph interpretations, there is a possibility of constructing sparse graph-based coding schemes and decoding over the joint graphical model representing the channels as well as the codes, as was done for joint detection and decoding of LDPC codes on partial response channels \cite{kur_02_tit_jid}. Instead of trying to build codes for the SDRC $\mathbf{P}_n$, the problem can be reduced to designing good codes and efficient decoding schemes for the FSCs $\mathbf{P}_{n, m}^\star$ for small values of $m$. For small deletion-replication probabilities $p$, which is the case of interest in practice, we can expect these codes to perform well for the SDRC $\mathbf{P}_n$ as well.

\section{Conclusions} \label{sec_conc}
We introduced a new channel model for a class of SECs which formulated the SEC as a channel with states. This allowed us to obtain analytical lower bounds for the capacity of SECs with only deletions or only replications. For the case of the BDC, we were able to write the SIR in terms of subsequence weights of binary sequences. Subsequence weights are known to be a quantity of interest in the maximum-likelihood decoding of sequences for the BDC (cf. Equation \eqref{eq_him}). Moreover, it is clear from Equation \eqref{eq_him} that the dependence of information rates for the BDC on the input statistics only appears in the term $\mathfrak{H}_m^{(i)}$, whereas the subsequence weights influence $\mathsf{H}(\overline{x})$ independently of the input statistics. Thus, our result establishes a natural link between the capacity of the BDC and the metric relevant for ML decoding. We were also able to obtain lower bounds on the capacity of the BDC that are known to be tight for small deletion probabilities. For the BRC, we were able to exactly characterize the Markov-$1$ rate, which {\textcolor{black}{agrees} 
with the \textcolor{black}{rates evaluated numerically} 
in \cite{mer_12_tit_nsc}, and with the expression in \cite{dri_07_tit_implbddc} which was proven with a completely different proof technique.}

For the case of an SEC with deletions and replications, we were able to provide a sequence of approximating FSCs that are totally ordered with respect to the mutual information rates achievable, and therefore, with respect to capacities. These approximating FSCs were shown to be such that the mutual information rate achievable for the SEC was equal to the limit of the mutual information rates achievable for the sequence of FSCs. To obtain numerical estimates of achievable rates on the DRC, we defined another sequence of indecomposable FSCs. Computing the mutual information rates for this sequence of FSCs allows us to relate the mutual information rate for the DRC to the limiting value of the mutual information rates of the sequence. For the particular case of the SDRC, we were able to show a stronger form of convergence of these mutual information rates.

The formulation in this paper not only allows us to get estimates of mutual information rates achievable on SECs but also gives some insight into possible code constructions and decoding schemes for such channels. The approximations introduced for the DRC gives us a natural way to reduce these problems. One would therefore obtain progressively better performing codes for the DRC by designing good codes for the sequence of approximating FSCs. We expect that for small values of the deletion-replication probability, a code constructed for an approximation with a moderate value of $m$ will perform well over the DRC as well. Some coding schemes for special cases of the FSCs (with $m = 1$) have been known in various contexts (See Remark \ref{rem_gen} and references therein
). Extending these schemes to better approximations (larger $m$ values) will prove crucial in designing good codes for the DRC. We emphasize that although the present paper considers only \emph{binary} SECs, the results extend naturally to the case of larger finite alphabets. The expressions for information rates will perhaps become more complicated, but the methods to arrive at their bounds or numerical estimates remain unchanged.

\appendices

\section{Proof of Proposition \ref{prop_chnequiv}} \label{app_chnequiv}
Both $\mathbf{Q}_n$ and $\mathbf{P}_n$ have the same input and output alphabets $\mathbb{X}$ and $\mathbb{Y}$, respectively. {That the} transition probabilities $Q_n$ and $P_n$ in Equations \eqref{eq_msectp} and {\eqref{eq_eqchntp}} are identical is evident by the following observations:
\begin{enumerate}[(i)]
\item \label{cons_zfromy} {Given the input sequence $x_{[n]}$}, for every parsing of $\overline{y} \in \overline{\mbb{Y}}$ as $\overline{y}_{[n]}$ in Equation \eqref{eq_msectp}, there is a corresponding {primary compatible} state path ${z_{[|\overline{y}|]}} \in {\mathbb{Z}^{|\overline{y}|}}$ in Equation {\eqref{eq_eqchntp}}. {Let us construct this state path for the case when the input $x_{[n]}$ is transformed to the output $\overline{y}_{[n]}$ where $\overline{y}_i = x_i^{\ell_i}$, i.e., the output substring $\overline{y}_i$ is the input symbol $x_i$ repeated $\ell_i \geq 0$ times. Here we write $x^0 = \lambda$. Note that there is no loss of generality here since $\overline{y}_i = x^{\ell_i}_i$ corresponds to every possbile output produced with non-zero probability by the DRC in Example \ref{eg_ddc}. }
{We first compute the total length of the output sequence $\ell = \sum_{i = 1}^n\ell_i = |\overline{y}|$. We then construct the $\Gamma_{[\ell]}$ sequence $\gamma_{[\ell]}$ as follows: for each $i \in [\ell]$, we set $\gamma_i = j \in [n]$ if $y_i$ is a part of the substring $\overline{y}_j$. Here $y_i \in \mathbb{Y} = \mathbb{X}$, or in other words, $y_i \neq \lambda$. By definition of the DRC, the \emph{order} of the input symbols is preserved so that the sequence $\gamma_{[\ell]}$ will be non-decreasing. We can therefore obtain the 
compatible path $z_{[\ell]}$ from $\gamma_{[\ell]}$.}
\item For every compatible state path {$z_{[\ell]} \in Z^\ell, \ell \geq 1$} in Equation {\eqref{eq_eqchntp}}, {given the input sequence $x_{[n]}$,} there is a corresponding parsing ${\overline{y}_{[n]} \in \mathbb{Y}^\ell}$ in Equation \eqref{eq_msectp}. {We construct this parsing as follows. We first 
obtain the sequence $\gamma_{[\ell]}$ corresponding to $z_{[\ell]}$. For each $i \in [n]$, set $\overline{y}_i = x_i^{w_i(\gamma_{[\ell]})}$, where $w_a(b)$ denotes the number of occurrences of $a$'s in the sequence $b$.} \textcolor{black}{Also, following the procedure in \eqref{cons_zfromy} above for the so constructed $\overline{y}_{[n]}$, it is easy to confirm that we end up with the same $z_{[\ell]}$ that we started with, so that the transformation between the parsed string $\overline{y}_{[n]}$ and the compatible state path $z_{[\ell]}$ is invertible.}
\item {Let us denote the 
compatible state path $z_{[\ell]}$ corresponding to parsing $\overline{y}_{[n]}$ as $z_{[\ell]}(\overline{y}_{[n]})$. Clearly, we have $\ell = |\circ(\overline{y}_{[n]})|$.} 
{Note that $i - z_i(\overline{y}_{[n]}) = \gamma_i(\overline{y}_{[n]}) = j$ (See \eqref{cons_zfromy} above) if $y_i$ is a part of the substring $\overline{y}_j = x_j^{\ell_j}$, i.e., $y_i = x_j$. This is true because $\ell_j > 0$ since a symbol $y_i$ other than $\lambda$ is produced at the channel output. 
We only need to prove}
\begin{equation} \label{eq_toproveeq}
{q_n(\overline{y}_{[n]}\mid x_{[n]}) = \p{d}^{n - \ell + z_{\ell}(\overline{y}_{[n]})}\prod_{i = 1}^\ell\msf{P}(Z_i = z_i(\overline{y}_{[n]})\mid Z_{i - 1} = z_{i - 1}(\overline{y}_{[n]})).}
\end{equation}
\textcolor{black}{This is a straightforward proof by induction and is omitted here.}
\end{enumerate}

\section{Special Cases of $D_i^\mathrm{iud}$} \label{app_splli}
\subsection{Bounds for $D_2^\mathrm{iud}$} \label{ssec_i2}
It is easy to see that when $i = 2$, Equation \eqref{eq_him} reduces to
\begin{align}
\mathfrak{H}^{(2)}_m &= 
\log_2(m + 1) - \frac{1}{2^{m + 1}}\sum_{j = 0}^{m + 1}{m + 1\choose j}h_2\Big(\frac{j}{m + 1}\Big), \label{eq_h2m}
\end{align}
%
\textcolor{black}{so we can bound} $\mathfrak{H}^{(2)}_m \geq \log_2(m + 1) - 1 + 2^{-m}$. This gives us \textcolor{black}{the desired bound}
\begin{align}
D_2^{\mrm{iud}} &\geq (1 - p) ^ 3\Big(\frac{4}{(2 - p) ^ 2} + \sum_{m \geq 0}(m + 1)p^m\log_2(m + 1)\Big) 
- h_2(p). 
\end{align}
It turns out that for
\begin{equation}
p < p^* \triangleq \exp\Big(-\frac{1 + \ln 2}{2\ln 2}\Big) \approx 0.294832606,
\end{equation}
we can \textcolor{black}{further} lower bound 
\begin{align}
&D_2^\mathrm{iud} \geq \frac{4(1 - p)^3}{(2 - p)^2} - h_2(p) \\
&\quad + (1 - p)^3\frac{\log_2e}{\ln p}\Big(\frac{p}{\ln p}(1 + \ln 2) - 2p\ln 2 - \frac{1}{p}\mathrm{Ei}(2\ln p)\Big),
\end{align}
\textcolor{black}{where $\mathrm{Ei}(x)$ is the \emph{exponential integral} function defined as
$\mathrm{Ei}(x) = \int_{-\infty}^x\frac{e^t}{t}dt$,
which can be numerically evaluated to arbitrary accuracy through a Taylor series expansion. However, }
%
this \textcolor{black}{leads to a} small-$p$ series expansion of $D_2^\mathrm{iud}$ \textcolor{black}{that} is no better than that of the lower bound for the BDC in 
\eqref{eq_capbndsspl}. We will improve this bound for small $p$ in the next subsection.

\subsection{Bounds for the case when $m = 1$} \label{ssec_m1}
We now pursue the other case where \eqref{eq_him} is easy to evaluate. Instead of evaluating $D_i^{\mrm{iud}}$ exactly, we can further lower bound it as follows.
\begin{align}
D_i^{\mrm{iud}} 
&\geq 1 - p - h_2(p) + (1 - p) \notag \\
&\quad \times \Big(\sum_{m = 0}^j \msf{P}(Z_i = -m \mid Z_0 = 0) H(Z_1 \mid Z_0 = 0, Z_i = -m, \mcl{X}, \mcl{Y})\Big) \\
&\triangleq 1 - p - h_2(p) + (1 - p)\Psi^{(i)}_j \\
&\triangleq \mathfrak{D}^{(i)}_j{\ }\forall{\ }j \geq 0, i \geq 1. 
\end{align}
We are essentially writing a series expansion for $D_i^\mathrm{iud}$ and lower bounding\footnote{All terms in the series expansion are non-negative.} it by the $j^\text{th}$ partial sum. Note that we can write
\begin{align}
\Psi^{(i)}_j 
&= \Psi_{j - 1}^{(i)} + \psi_{i, j}p^j(1 - p)^i \label{eq_aterm}
\end{align}
where $\psi_{i, m}$ was defined in Theorem \ref{thm_sirbdc}. Clearly, the sequence $\{\Psi_j^{(i)}\}_{j \geq 0}$ is non-decreasing, and, in turn, so is the sequence $\{\mathfrak{D}^{(i)}_j\}_{j \geq 0}$. Since $\Psi^{(i)}_0 = \psi_{i, 0} = 0$, we have $\Psi^{(i)}_1 = p(1 - p)^i\psi_{i, 1}$. Further, by definition, 
$D_i^\mathrm{iud} = \lim_{j \rightarrow \infty}\mathfrak{D}^{(i)}_j = \sup_{j \geq 0}\mathfrak{D}^{(i)}_j$. 
Thus for every $j \geq 0$, we can write
\begin{align}
C^\mathrm{iud}_\mathrm{BDC} &= \sup_{i \geq 1} D_i^{\mrm{iud}} 
\geq \sup_{i \geq 1}\mathfrak{D}^{(i)}_j \triangleq \mathfrak{D}^{\mrm{iud}}_j
.
\end{align}

From the channel model, $\msf{P}(X_{[i]} = x_{[i]} \mid Z_0 = 0, Z_i = -1) = 2^{-i}$ since $\mcl{X} \ind \mcl{Z}$ and $\mcl{X}$ is i.u.d., and
\begin{equation}
\msf{P}(Y_{[i - 1]} = y_{[i - 1]} \mid X_{[i]} = x_{[i]}, Z_0 = 0, Z_i = -1) = \frac{w_{y_{[i - 1]}}(x_{[i]})}{i}.
\end{equation}
For $y_{[i - 1]} = x_{[i - 1] - z_{[i - 1]}}$ for some realization $z_{[i - 1]}$ with the boundary conditions $z_0 = 0$ and $z_i = -1$,
\begin{align}
H(Z_1 \mid Z_i = -1, &X_{[i]} = x_{[i]}, Y_{[i - 1]} = y_{[i - 1]}) \\
&= h_2\Big(\frac{1}{r_1(x_{[i]})}\Big)\mathds{1}_{\mcl{R}_1(x_{[i]}, y_{[i - 1]})} 
\end{align}
where $\mcl{R}_1(x_{[i]}, y_{[i - 1]})$ is the event that the single deletion occurred in the first run of $x_{[i]}$ to result in $y_{[i - 1]}$. To see this, let $y_{[i - 1]}$ represent a received word resulting from a single deletion upon transmission of $x_{[i]}$. Consider the two mutually exclusive and exhaustive cases in this scenario:
\begin{itemize}
\item The single deletion occurs in a run other than the first run of $x_{[i]}$. In this case, there is no ambiguity that $Z_1 = 0$, and the first run of $y_{[i - 1]}$ is either the same or larger than\footnote{When the second run of $x_{[i]}$ disappears.} that of $x_{[i]}$.
\item The single deletion occurs in the first run of $x_{[i]}$.
\begin{itemize}
\item If $r_1(x_{[i]}) = 1$, there is no ambiguity that $Z_1 = -1$.
\item If $r_1(x_{[i]}) > 1$, the deleted symbol could be, with equal likelihood, one of the symbols comprising the first run of $x_{[i]}$. The uncertainty in $Z_1$ is $h_2\Big(\frac{1}{r_1(x_{[i]})}\Big)$.
\end{itemize}
In both the above cases, the uncertainty can be written as $h_2\Big(\frac{1}{r_1(x_{[i]})}\Big)$.
\end{itemize}
Therefore, 
\begin{align}
\psi_{i, 1} &= i\sum_{x_{[i]}}\frac{1}{2^i}\sum_{y_{[i - 1]}}\frac{w_{y_{[i - 1]}}(x_{[i]})}{i}h_2\Big(\frac{1}{r_1(x_{[i]})}\Big)\mds{1}_{\mcl{R}_1(x_{[i]}, y_{[i - 1]})} \\
&= i\sum_{x_{[i]}}\frac{1}{2^i}\frac{r_1(x_{[i]})}{i}h_2\Big(\frac{1}{r_1(x_{[i]})}\Big) \\
&= \frac{1}{2}\sum_{j = 1}^{i - 2} \frac{j}{2^j}\log_2 j + 2\frac{i}{2^i}\log_2i. \label{eq_aiterm} 
\end{align}
We observe that $\psi_{i, 1}$ is non-decreasing in $i$, and converges exponentially to the value
$\psi_1 \approx 1.288531275$. 
From \eqref{eq_aterm} and \eqref{eq_aiterm}, we have
$
D^{\mrm{iud}}_i = 1 + p\log_2p - p\log_2(2e) + \psi_{i, 1}p + O(p^2)$
and from Equation \eqref{eq_sirbdc},
\begin{equation} \label{eq_ciudsdp}
C^{\mathrm{iud}}_\mathrm{BDC} = 1 + p\log_2p - \mathtt{d}p + O(p^2)
\end{equation}
where $\mathtt{d} = \log_2(2e) - \psi_1 \approx 1.154163765$. We note here that this is exactly the same bound obtained in \cite{kan_10_isit_delchn} with a completely different technique. Since this bound was shown to be tight for small $p$, we have that the capacity of the BDC itself is given by the above expression for small $p$.

\begin{dis} \vspace*{2mm}
The advantage in the evaluation of the above bound was that, when we restrict to the case of a single deletion, the ambiguity in the first channel state $Z_1$ arises only when $r_1(x_{[i]}) > 1$, in which case the uncertainty is exactly $h_2\Big(\frac{1}{r_1(x_{[i]})}\Big)$. This, however, is not true when there are $2$ or more deletions. 
\textcolor{black}{Although similar methods can be used to arrive at the corresponding bounds for first-order Markov sources, these bounds are not improved much over their SIR counterparts.}
\end{dis}

\section{Proofs of Results for the BRC} \label{app_brc}
\subsection{Proof of Corollary \ref{cor_brclb}} \label{ssec_brclb}
We have from 
Lemma \ref{lem_monri} that
\begin{align}
C^{\mathcal{M}1}_{\mathrm{BRC}} \geq R_2^{\mathcal{M}1} \triangleq &\max_{\alpha} \Big[ h_2(\alpha) + \frac{H(Z_1 \mid Z_0 = 0, Z_2, {X_0}, \mathcal{X}, \mathcal{Y})}{1 - p} \notag \\
&\qquad- \frac{p + (1 - \alpha)(1 - p)}{1 - p}h_2\Big(\frac{p}{p + (1 - \alpha)(1 - p)}\Big)\Big] 
\end{align}
where we have used the expression for $H(Z_1 \mid Z_0 = 0, {Y_0}, \mathcal{Y})$ from the proof of Theorem \ref{thm_m1rbrc}. Observe that $Z_2 \in \{0, 1, 2\}$, and among these possibilities, the only event wherein there is an ambiguity in the value of $Z_1$ is when $Z_2 = 1$. Thus, we can see easily that $H(Z_1 \mid Z_0 = 0, Z_2, \mathcal{X}, \mathcal{Y}) = 2p(1 - p)(1 - \alpha)$. Hence
\begin{align}
R_2^{\mathcal{M}1} &= \max_{\alpha} \Big[ h_2(\alpha) + 2p(1 - \alpha) 
- \frac{p + (1 - \alpha)(1 - p)}{1 - p}h_2\Big(\frac{p}{p + (1 - \alpha)(1 - p)}\Big)\Big]. 
\end{align}
It can be shown that the optimal $\alpha$ in the above is given by
\begin{equation}
\alpha^* = \frac{1}{(1 - p)(2^{2p} + 1)}.
\end{equation}
Note that $\alpha^*$ is always larger than $\frac{1}{2}$, and $\alpha^* \leq 1$ for $p \leq p_*$ where
$p_* \approx 0.734675821$.
 Plugging this back in the expression for $R_2^{\mathcal{M}1}$ ends the proof.

\subsection{Proof of Corollary \ref{cor_srpbrc}} \label{ssec_srpbrc}
From 
Lemma \ref{lem_monri}, for i.u.d. inputs,
\begin{align}
C^\mathrm{iud}_\mathrm{BRC} &= 1 - \frac{H(Z_1 \mid Z_0 = 0, {Y_0}, \mathcal{Y})}{1 - p} + \sup_{i \geq 1} \frac{H(Z_1 \mid Z_0 = 0, Z_i, {X_0}, \mcl{X}, \mcl{Y})}{1 - p} \\
&= 1 - \frac{1 + p}{2(1 - p)}h_2\Big(\frac{2p}{1 + p}\Big) 
+ \sup_{i \geq 1}\Big(ip(1 - p)^{i - 2}H(Z_1 \mid Z_0 = 0, Z_i = 1, {X_0}, \mcl{X}, \mcl{Y})\Big) + O(p^2), 
\end{align}
where we have used the expression for $H(Z_1 \mid Z_0 = 0, {Y_0}, \mathcal{Y})$ from the proof of Theorem \ref{thm_m1rbrc} for $\alpha = \frac{1}{2}$. The last equality is true since $H(Z_1 \mid Z_0 = 0, Z_i = 0, {X_0}, \mathcal{X}, \mathcal{Y}) = 0$. 
As shown in the proof of Theorem \ref{thm_m1rbrc}, we can write
\begin{align}
H(Z_1 \mid Z_0 = 0, Z_i = 1, 
{X_0}, \mathcal{X}, \mathcal{Y}) 
&= \mathsf{E}[h_2(\mathsf{P}(Z_1 = 0 \mid Z_0 = 0, Z_i = 1, {x_0}, r_1^0(x_\mathbb{N}), r_1^0(y_\mathbb{N})))]. 
\end{align}
Further, there is no ambiguity in $Z_1$ if the single replication does not occur in the first run of $x_\mathbb{N}$. Therefore, for a first-order Markov input process,
\begin{align}
H(Z_1 
\mid Z_0 = 0, Z_i = 1, {X_0}, \mcl{X}, \mcl{Y}) 
&
= \mathsf{E}[h_2(\mathsf{P}(Z_1 = 0\mid Z_0 = 0, Z_i = 1, 
 {x_0}, r_1^0(x_\mathbb{N}) = l, r_1^0(y_\mathbb{N}) = l + 1))] \notag \\
&= \sum_{l = 1}^{i - 1}(1 - \alpha)^l\alpha\frac{l + 1}{i}h_2\Big(\frac{1}{l + 1}\Big) + (1 - \alpha)^ih_2\Big(\frac{1}{i}\Big). \notag
\end{align}
For $\alpha = \frac{1}{2}$, we get 
$iH(Z_1\mid Z_i = 1, {X_0}, \mathcal{X}, \mathcal{Y}) 
= \frac{1}{2}\sum_{l = 1}^{i - 2}\frac{l}{2^l}\log_2l + 2\frac{i}{2^i}\log_2i = \psi_{i, 1}$ 
from Equation\eqref{eq_aiterm}, and therefore
\begin{align}
C^{\mrm{iud}}_\mathrm{BRC} &= 1 - \frac{1 + p}{2(1 - p)}h_2\Big(\frac{2p}{1 + p}\Big) 
+ \sup_{i \geq 1}\Big(p(1 - p)^{i - 2}\psi_{i, 1}\Big) + O(p^2), \\
&= 1 + p\log_2p + \mathtt{r}p + O(p^2), 
\end{align}
where $\mathtt{r} = \log_2(\frac{2}{e}) + \psi_1 = 2 - \mathtt{d} \approx 0.845836235$. As was the case for the BDC, we expect this to be a tight bound for the capacity for small $p$.

\section{Extending the Measures $\mathsf{P}_{\langle m\rangle}$ to $\mathscr{B}$} \label{app_extmeas}
By defining the 
transition probabilities $\mathsf{P}_{\langle m\rangle}(Z_i^{(m)} \mid Z_{i - 1}^{(m)})$ as in Section \ref{ssec_appsc}, the measures $\mathsf{P}_{\langle m\rangle}$ are well-defined over $\mathscr{G}_m = \sigma(\mathcal{Z}^{(m)})$. Let $(\mathcal{Z}^{(m)})^{-1}(\overline{z}) \triangleq \{\varsigma \in \mathbb{S}_\mathcal{Z} : \mathcal{Z}^{(m)}(\varsigma) = \overline{z}\}$ for $\overline{z} \in \overline{\mathbb{Z}}_{\pm m}$, and similarly $\mathcal{Z}^{-1}(\overline{z}) \triangleq \{\varsigma \in \mathbb{S}_\mathcal{Z} : \mathcal{Z}(\varsigma) = \overline{z}\}$. Then, clearly
\begin{equation}
\mathcal{Z}^{-1}(\overline{z}) \subset (\mathcal{Z}^{(m)})^{-1}(\overline{z}){\ }\forall{\ }\overline{z} \in \overline{\mathbb{Z}}_{\pm m}
\end{equation}
and
\begin{equation}
\mathcal{Z}^{-1}(\overline{z}) \in \mathscr{B}_\mathcal{Z}, (\mathcal{Z}^{(m)})^{-1}(\overline{z}) \in \mathscr{G}_m{\ }\forall{\ }\overline{z} \in \overline{\mathbb{Z}}_{\pm m}.
\end{equation}
Then, we define
\begin{equation} \label{eq_pscomp}
\mathsf{P}_{\langle m\rangle}(\mathcal{Z}^{-1}(\overline{z})) = \mathsf{P}_{\langle m\rangle}((\mathcal{Z}^{(m)})^{-1}(\overline{z})){\ }\forall{\ }\overline{z} \in \overline{\mathbb{Z}}_{\pm m}.
\end{equation}
This will imply that for every $\overline{z} \in \overline{\mathbb{Z}}_{\pm m}$,
\begin{equation}
\mathsf{P}_{\langle m\rangle}((\mathcal{Z}^{(m)})^{-1}(\overline{z}) \setminus \mathcal{Z}^{-1}(\overline{z})) = 0.
\end{equation}
By definition, we also have for $\overline{z} \in \overline{\mathbb{Z}} \setminus \overline{\mathbb{Z}}_{\pm m}$ that $(\mathcal{Z}^{(m)})^{-1}(\overline{z}) = \emptyset$ so that the associated probability is zero under any measure $\mathsf{P}, \mathsf{P}_{\langle m\rangle}$. We can now consider the space $(\mathbb{S}_\mathcal{Z}, \mathscr{B}_\mathcal{Z}, \mathsf{P}_{\langle m\rangle})$ to be obtained from $(\mathbb{S}_\mathcal{Z}, \mathscr{G}_m, \mathsf{P}_{\langle m\rangle})$ along with the definition \eqref{eq_pscomp} and subsequent \emph{completion} \cite[\S 2.6.19]{res_05_bok_probpath}.

By now defining $\mathsf{P}_{\langle m\rangle}(\mathcal{X}) = \mathsf{P}(\mathcal{X})$ independent of $m$, we can extend the measure $\mathsf{P}_{\langle m\rangle}$ to $\mathscr{B} = \sigma(\{\mathcal{X}, \mathcal{Z}\})$ for each $m \in \mathbb{Z}^+$ as required.

\section{Proof of Proposition \ref{prop_sdrcap}} \label{app_sdrclim}
We start with two Lemmas that will be useful in proving this result.
\begin{lem} \label{lem_prub} \vspace*{2mm}
For the SDRC, for every $n \in \mathbb{N}$, let $m \in \mathbb{N}$. Then,
\begin{align}
\mathsf{P}_{\langle m\rangle}\Big(&\max_{i = 1}^{N_n^{(m)}} |Z_i| \geq m \textcolor{black}{\mid \mathfrak{B}_0^{(m)}}\Big) 
= \mathsf{P}\Big(\max_{i = 1}^{N_n^{(m)}} |Z_i| \geq m \textcolor{black}{\mid \mathfrak{B}_0}\Big) = O\Big(\frac{n + m}{m^2}\Big).\tag*{$\blacksquare$}
\end{align}
\end{lem}
\begin{IEEEproof}
\textcolor{black}{We first note that by definition of the measure $\mathsf{P}_{\langle m\rangle}$}
\begin{align}
\textcolor{black}{\mathsf{P}_{\langle m\rangle}\Big(\vartheta \in \mathbb{S}} &\textcolor{black}{: \max_{i = 1}^{N_n^{(m)}(\vartheta)}|Z_i(\vartheta)| < m, Z_0^{(m)}(\vartheta) = 0, N_0^{(m)}(\vartheta) = 0\Big)} \notag \\
&\textcolor{black}{= \mathsf{P}\Big(\vartheta \in \mathbb{S} : \max_{i = 1}^{N_n^{(m)}(\vartheta)}|Z_i(\vartheta)| < m, Z_0(\vartheta) = 0, N_0(\vartheta) = 0\Big).} 
\end{align}
\textcolor{black}{From this and the fact that $\mathsf{P}_{\langle m\rangle}(\mathfrak{B}_0^{(m)}) = \mathsf{P}(\mathfrak{B}_0)$, we have}
\begin{align}
\mathsf{P}_{\langle m\rangle}\Big(&\max_{i = 1}^{N_n^{(m)}}|Z_i| \geq m \mid \mathfrak{B}_0^{(m)}\Big) \notag \\
&= \mathsf{P}\Big(\max_{i = 1}^{N_n^{(m)}}|Z_i| \geq m \mid \mathfrak{B}_0\Big). \label{eq_step1}
\end{align}
\textcolor{black}{From the discussion following Lemma \ref{lem_indec}, we have $N_n^{(m)} \leq n + m$, which implies that we can bound}
\begin{equation}
\mathsf{P}\Big(\max_{i = 1}^{N_n^{(m)}} |Z_i| \geq m \mid \mathfrak{B}_0\Big) \leq \mathsf{P}(\max_{i = 1}^{n + m} |Z_i| \geq m \mid \mathfrak{B}_0) \textcolor{black}{ \leq \Big(\frac{1}{1-p}\Big)\cdot\mathsf{P}(\max_{i = 1}^{n + m} |Z_i| \geq m\mid\mathfrak{B}^\prime_0)}, \label{eq_step2}
\end{equation}
\textcolor{black}{where $\mathfrak{B}^\prime_0 \triangleq \{Z_0 = 0\}$. The last inequality is true because we have $\mathsf{P}(\cdot\mid\mathfrak{B}_0) = \frac{1}{1 - \p{r}}\mathsf{P}(\cdot\mid\mathfrak{B}^\prime_0)$ for $\vartheta \in \mathbb{S}$ such that $Z_1(\vartheta) \leq 0$, and for $\vartheta \in \mathbb{S}$ such that $Z_1(\vartheta) = 1$, the right hand side dominates trivially. Under the boundary condition $\mathfrak{B}_0^\prime$, it is easy to see that}
%
$Z_n$ 
\textcolor{black}{can be written as} the $n^\text{th}$ partial sum of the i.i.d. process $\{\Xi_i\}_{i \geq 1}$ \textcolor{black}{where}
\begin{equation}
\mathsf{P}(\Xi_1 = \xi) = 
\begin{cases}
\p{r}, &\xi = 1\\
\p{d}^{-\xi}\p{t}, &\xi \leq 0.
\end{cases}
\end{equation}
For the SDRC, we have $\mathsf{E}[\Xi_1] = \chi = 0$ and $\mathsf{Var}[\Xi_1] = \nu^2 = \frac{2p}{1 - p} < \infty$ since $p \in [0, 1)$. Hence, $Z_n \in L^2(\mathbb{S}, \mathscr{B}, \mathsf{P})$ for every $n \in \mathbb{N}$.

Let $\mathscr{S}_n = \sigma(\{Z_n\}) \subset \mathscr{B}$, the sigma-algebra generated by $Z_n$, for every $n \in \mathbb{N}$. Clearly, $\mathscr{S}_n = \sigma(\{\Xi_{[n]}\})$ so that $\{\mathscr{S}_n\}_{n \geq 1}$ is a filtration, and $Z_n \in \mathscr{S}_n$ by definition. Let $\mathscr{S}_n \uparrow \mathscr{S} \subset \mathscr{B}$ as $n \rightarrow \infty$. 
Since $\mathsf{E}[Z_n\mid \mathscr{S}_{n - 1}] = \mathsf{E}[Z_{n - 1} + \Xi_n\mid \mathscr{S}_{n - 1}] = Z_{n - 1}$, 
$\{Z_n, \mathscr{S}_n\}_{n \geq 1}$ is a martingale under the measure $\mathsf{P}(\cdot\mid\mathfrak{B}_0^\prime)$. Consequently, $\{|Z_n|, \mathscr{S}_n\}_{n \geq 1}$ is a submartingale. 
Since $|Z_n| \in L^2(\mathbb{S}, \mathscr{B}, \mathsf{P})$, from Doob's submartingale inequality \cite[\S 14.6]{wil_91_bok_probmart}, we have
\begin{equation}
\mathsf{P}(\max_{i = 1}^{n + m} |Z_i| \geq m \mid \mathfrak{B}_0^\prime) \leq \frac{\mathsf{E}[|Z_{n + m}|^2]}{m^2} = \Big(\frac{2p}{1 - p}\Big)\frac{n + m}{m^2}. \label{eq_step3}
\end{equation}
\textcolor{black}{Putting \eqref{eq_step1}, \eqref{eq_step2} and \eqref{eq_step3} together, we have the desired result.}
\end{IEEEproof}
\begin{lem} \label{lem_wc} \vspace*{2mm}
Let $(\mathbb{T}, \mathscr{A})$ be a measurable space, and let $\{\mathsf{Q}_n\}_{n \geq 1}$, $\mathsf{Q}$ all be probability measures on this space. Suppose that
\begin{enumerate}[i)]
\item For every $n \geq 1$, there is a set $\mathbb{B}_n \in \mathscr{A}$ such that $\mathsf{Q}_n(\mathbb{A}) = \mathsf{Q}(\mathbb{A})$ for every $\mathbb{A} \subset \mathbb{B}_n$, $\mathbb{A} \in \mathscr{A}$.
\item $\mathsf{Q}(\mathbb{B}_n) \rightarrow 1$ as $n \rightarrow \infty$.
\end{enumerate}
Then the measures $\mathsf{Q}_n$ converge in total variation to $\mathsf{Q}$, i.e., $\mathsf{Q}_n \stackrel{tv}{\longrightarrow} \mathsf{Q}$ as $n \rightarrow \infty$.
\end{lem}
\begin{IEEEproof}
From ii), for every $\epsilon > 0$, there exists $n^\prime(\epsilon) \in \mathbb{N}$ such that
\[
\mathsf{Q}(\mathbb{B}_n) \geq 1 - \epsilon{\ }\forall{\ }n \geq n^\prime(\epsilon).
\]
From i), $\mathsf{Q}_n(\mathbb{A} \cap \mathbb{B}_n) = \mathsf{Q}(\mathbb{A} \cap \mathbb{B}_n)$ for every $n \geq 1$, $\mathbb{A} \in \mathscr{A}$. Therefore, for every $\epsilon > 0$,
\begin{align}
||\mathsf{Q}_n - \mathsf{Q}|| &= 2\sup_{\mathbb{A} \in \mathscr{A}}|\mathsf{Q}_n(\mathbb{A}) - \mathsf{Q}(\mathbb{A})| \notag \\
&= 2\sup_{\mathbb{A} \in \mathscr{A}}|\mathsf{Q}_n(\mathbb{A} \cap \mathbb{B}_n^\mathsf{C}) - \mathsf{Q}(\mathbb{A} \cap \mathbb{B}_n^\mathsf{C})| \notag \\
&\leq 2\epsilon{\ }\forall{\ }n \geq n^\prime(\epsilon). \notag
\end{align}
Hence $\mathsf{Q}_n \stackrel{tv}{\longrightarrow} \mathsf{Q}$ as $n \rightarrow \infty$.\vspace*{2mm}
\end{IEEEproof}

Note that 
$
\mathbb{D}_{n, m} \triangleq \Big\{\vartheta \in \mathbb{S} : \max_{i = 1}^{N_n^{(m)}(\vartheta)}|Z_i(\vartheta)| \geq m\Big\}
$
 is the subset of $\mathbb{S}$ in $\mathscr{B}$ where $\mathsf{P}_{\langle m\rangle}(X_{[n]}, Y_{[N_n^{(m)}]}^{(m)}\mid\mathfrak{B}_0^{(m)})$ differs from $\mathsf{P}(X_{[n]}, Y_{[N_n]}\mid\mathfrak{B}_0)$. From Lemma \ref{lem_prub}, we have
\[
\mathsf{P}_{\langle m(n)\rangle}(\mathbb{D}_{n, m(n)}\textcolor{black}{ \mid \mathfrak{B}_0^{(m(n))}}) = \mathsf{P}(\mathbb{D}_{n, m(n)} \textcolor{black}{ \mid \mathfrak{B}_0}) \rightarrow 0
\]
as $n \rightarrow \infty$, whenever $m(n) = \omega(\sqrt{n})$. 
\textcolor{black}{By setting $\mathbb{T} = \mathbb{S}$, $\mathscr{A} = \mathscr{B}$, $\mathsf{Q} = \mathsf{P(\cdot\mid\mathfrak{B}_0)}$, and for each $n \in \mathbb{N}$, $\mathsf{Q}_n = \mathsf{P}_{\langle m(n)\rangle}(~\cdot~\mid~\mathfrak{B}_0^{(m(n))})$ and $\mathbb{B}_n = \mathbb{D}_{n, m(n)}^\mathsf{C}$, and extending $\mathsf{P}_{\langle m(n)\rangle}$ to $\mathscr{B}$ (cf. Appendix \ref{app_extmeas}) such that it agrees with the measure $\mathsf{P}$ on every subset of $\mathbb{B}_n$ for each $n \geq 1$,} 
both conditions i) and ii) in Lemma \ref{lem_wc} are satisfied. From this and \cite[Corollary 1$^\prime$]{dob_60_tproba_limexp}, we have the desired result.\vspace*{2mm}

\section*{Acknowledgment}
{The authors are very grateful to the anonymous reviewers and Hugues Mercier for their questions, comments and suggestions that have tremendously improved the presentation and rigor of the paper. They thank Hugues Mercier for bringing a recent publication with related results, \cite{mer_12_tit_nsc}, to their attention, and for providing bounds from the same paper that have been included in Figs. \ref{fig_bdc}, \ref{fig_brc} and \ref{fig_sdrccap}.} The authors also thank Eleni Drinea and Michael Mitzenmacher, respectively, for providing the numerical lower bounds for the BDC from \cite{kir_10_tit_ddccap} and the BRC from \cite{mit_08_tit_sticky} (that appear as black circles in Figs. \ref{fig_bdc} and \ref{fig_brc}, respectively). A.~R.~Iyengar would like to thank Aman Bhatia, Suhas Diggavi, Patrick Fitzsimmons, Henry Pfister, Bharath Sriperumbudur \& Ruth Williams for stimulating discussions.

\twobibs{
\bibliographystyle{IEEEtran}
\bibliography{../../../Bib/mybib}
}
{

}

\end{document}